\documentclass[journal=jacsat,manuscript=article]{achemso}
\usepackage{adjustbox}
\usepackage[colorlinks=true, allcolors=blue]{hyperref}
\usepackage{amsmath}
\usepackage{kmath}
\usepackage{amssymb}
\usepackage[version=4]{mhchem}
\usepackage{bbding}
\usepackage{setspace}
\usepackage{placeins}
\usepackage{pifont}
\usepackage{xcolor}
\usepackage[normalem]{ulem}
\usepackage{verbatim}
\usepackage{indentfirst}
\usepackage{tabularray}
\usepackage{textgreek}

\usepackage{rotating}
\usepackage{lipsum}

\SectionNumbersOn

\author{Dominik Lüthi}
\altaffiliation{These authors contributed equally.}
\affiliation[1]{Empa - Swiss Federal Laboratories for Materials Science and Technology, nanotech@surfaces Laboratory, Ueberlandstrasse 129, 8600 D\"ubendorf, Switzerland}
\alsoaffiliation[2]{Department of Chemistry, Biochemistry and Pharmaceutical Sciences, University of Bern, 3012 Bern, Switzerland}
\author{Lin Yang}
\altaffiliation{These authors contributed equally.}
\affiliation[3]{Max Planck Institute of Microstructure Physics, Weinberg 2, 06120 Halle, Germany}
\author{Xiuling Yu}
\affiliation[4]{Centre for Advancing Electronics Dresden (cfaed), Department of Chemistry and Food Chemistry, Technische Universität Dresden, 01062 Dresden, Germany}
\author{Ji Ma}
\affiliation[5]{College of Materials Science and Opto-Electronic Technology and Center of Materials Science and Optoelectronics Engineering, University of Chinese Academy of Science, 100049 Beijing, P. R. China}
\author{Xinliang Feng}
\affiliation[3]{Max Planck Institute of Microstructure Physics, Weinberg 2, 06120 Halle, Germany}
\alsoaffiliation[4]{Centre for Advancing Electronics Dresden (cfaed), Department of Chemistry and Food Chemistry, Technische Universität Dresden, 01062 Dresden, Germany}
\author{Carlo A. Pignedoli}
\affiliation[1]{Empa - Swiss Federal Laboratories for Materials Science and Technology, nanotech@surfaces Laboratory, Ueberlandstrasse 129, 8600 D\"ubendorf, Switzerland}
\author{Roman Fasel}
\affiliation[1]{Empa - Swiss Federal Laboratories for Materials Science and Technology, nanotech@surfaces Laboratory, Ueberlandstrasse 129, 8600 D\"ubendorf, Switzerland}
\alsoaffiliation[2]{Department of Chemistry, Biochemistry and Pharmaceutical Sciences, University of Bern, 3012 Bern, Switzerland}
\author{Gabriela Borin Barin}
\affiliation[1]{Empa - Swiss Federal Laboratories for Materials Science and Technology, nanotech@surfaces Laboratory, Ueberlandstrasse 129, 8600 D\"ubendorf, Switzerland}
\email{gabriela.borin-barin@empa.ch}
\title{Interfacial Coupling and Sparse Intercalation of 7-Atom-Wide Armchair Graphene Nanoribbons by N-Heterocyclic Carbene Monolayers}

\mciteErrorOnUnknownfalse

\begin{document}
\begin{abstract}
Graphene nanoribbons (GNRs) synthesized on metal substrates experience electronic coupling and screening from the underlying surface, which, although often weak, can modify their observed properties and complicate their transfer to device-compatible substrates. Intercalation of GNRs by self-assembled monolayers (SAMs) offers a possible route to reduce this interaction. Here, we investigate the intercalation of 7-atom-wide armchair graphene nanoribbons (7-AGNRs) on Au(111) using N-heterocyclic carbenes (NHCs). Low-temperature scanning tunneling microscopy and spectroscopy, Raman spectroscopy, and density functional theory calculations reveal that the adsorption geometry of the NHCs strongly influences the intercalation yield for GNRs. Methyl-substituted NHCs form flat-lying dimers that partially intercalate the GNRs, producing locally decoupled segments. In contrast, bulkier isopropyl-substituted NHCs form upright monomers that embed the GNRs within the monolayer, preventing intercalation. The low intercalation yield indicates that lifting the nanoribbon from the Au surface is energetically costly. These results establish molecular adsorption geometry and packing as key parameters controlling intercalation at GNR–metal interfaces, with implications for the rational design of decoupling layers for GNR-based device integration pathways.
\end{abstract}

\section*{Introduction}

Graphene nanoribbons (GNRs) are quasi-one-dimensional strips of graphene whose electronic and optical properties are governed by their width, edge configuration, and overall geometry.\cite{son2006energy, yazyev2013guide, yano2019quest} Atomically precise GNRs are synthesized through a bottom-up approach that combines surface-assisted polymerization of molecular precursors with a subsequent cyclodehydrogenation step, typically carried out on metallic substrates under ultra-high vacuum (UHV) conditions. \cite{cai2010atomically, talirz2017surface, ruffieux2016surface, kinikar2023surface} Rational design of the molecular precursors enables precise control over the final geometry of the ribbon and consequently its electronic structure\cite{narita2019solution}. A wide variety of GNRs—such as armchair (AGNRs)\cite{cai2010atomically, yamaguchi2020small, kimouche2015ultra, chen2013tuning, hwang2025optimized}, chevron-type\cite{cai2010atomically, geagea2024growth, cao2018tuning}, edge-extended\cite{kinikar2023surface, wang2021cove, groning2018engineering, xiang2025zigzag} and zigzag GNRs (ZGNRs)\cite{ruffieux2016surface, blackwell2021spin}—have been successfully synthesized using this approach\cite{zhang2026bottom}. 
Such structural tunability is essential for tailoring device functionality. It enables control over the band gap of AGNRs through their width\cite{yang2007quasiparticle, merino2017width}, while in other types of GNR it leads to the emergence of spin-polarized edge states, topological boundary states, and other exotic electronic properties that may be harnessed for spintronic applications.\cite{wang2016giant, ruffieux2016surface, brede2023detecting, rizzo2018topological}

Integrating atomically precise GNRs into devices, particularly on insulating substrates such as silicon dioxide, requires a high-yield, nondestructive substrate transfer step from the metallic growth surface. Although AGNRs are chemically stable and have been successfully transferred using several methods,\cite{overbeck2019optimized, richter2020charge, houtsma2021atomically} more reactive GNRs, such as those with edge extensions, zigzag edges, or other non-armchair geometries, interact strongly with the metallic substrate\cite{ruffieux2016surface, lawrence2022circumventing}, making high-quality transfer more difficult. This is especially problematic for GNRs that feature open-shell configurations or unpaired $\pi$-electrons at their edges, as these edge extensions or topologies are intrinsically reactive and prone to degradation\cite{lawrence2022circumventing, ruffieux2016surface}, making it impossible to expose them to ambient conditions or remove them from UHV.

Transfer methods such as Au/mica etching\cite{borin2019surface} or electrochemical delamination\cite{overbeck2019universal, senkovskiy2017making} are solution-based and therefore generally incompatible with reactive GNRs and UHV requirements. Moreover, solution transfer often results in contamination, wrinkles, and low yield,\cite{10.1039/c7nr07369k, richter2020charge, bae2010roll} 
which impairs reproducibility and device performance.\cite{10.1002/adfm.202103798} Recent STM studies confirmed this limitation, showing that while stable 9-AGNRs survive polymer-free wet transfer, more reactive GNRs with extended or open-shell edge topologies degrade or even disintegrate under such conditions.\cite{doi:10.1021/acsanm.5c02753} As a result, more sophisticated transfer techniques are required to preserve the electronic properties while successfully transferring GNRs to a final substrate beneficial for device fabrication\cite{zhang2026bottom}.

Intercalation offers a promising approach to decouple GNRs from their metallic growth substrates. By introducing atoms or molecules with a strong affinity for the metal surface, a self-assembled monolayer (SAM) can form beneath the GNRs, thereby reducing their interaction with the underlying substrate and enabling a cleaner and less invasive transfer.\cite{ohtomo2016etchant, 10.1021/acs.jpcc.5b00528} In addition to aiding transfer, intercalation can significantly alter the electronic environment at the interface\cite{kinikar2024electronic}. This may enable the observation of spectral features typically suppressed by the metallic substrate, such as frontier molecular orbitals and localized edge states.\cite{kinikar2024electronic, deniz2017revealing} Intercalation also enables systematic studies of how GNRs respond to substrates with modified physical or chemical properties. For example, intercalation can tune substrate work functions by nearly 2 eV in layered graphene structures, offering flexible electronic control,\cite{arnay2025tuning, 10.1088/1674-1056/ac6941, daukiya2018highly} and rare-earth intercalation of GNRs on TbAu\textsubscript{2}/Au(111) shifts their doping from p- to weak n-type while preserving the band gap, highlighting both work-function modulation and magnetic substrate engineering.\cite{que2020surface, edens2025spin} More generally, varying substrate properties such as charge transfer potential,\cite{deniz2017revealing} dielectric screening,\cite{el2020controlled} or symmetry can strongly influence the electronic structure of GNRs, providing controlled routes to probe substrate–nanoribbon interactions.

In this work, we explore the intercalation of 7-atom-wide AGNRs (7-AGNRs) with N-heterocyclic carbenes (NHCs) --- a class of molecules known for their strong covalent bonding with gold surfaces under UHV conditions.\cite{crudden2014ultrastable, larrea2017, inayeh2021self} NHCs can form robust assemblies without requiring thermal activation, making them particularly attractive for systems like GNRs, which can be sensitive to elevated temperatures during processing.\cite{larrea2017, zhukhovitskiy2015carbene, engel2017new, kaur2022fundamentals} 
We focus on two specific NHC derivatives: the methyl-substituted NHC (NHC\textsuperscript{Me})\cite{larrea2017} and the bulkier isopropyl-substituted NHC (NHC\textsuperscript{\textit{i}Pr})\cite{bakker2018elucidating}, which exhibit adsorption geometries different from each other on Au(111).\cite{inayeh2021self}
Using scanning tunneling microscopy and spectroscopy (STM/STS) and Raman spectroscopy, we investigate the morphological and spectroscopic signatures of intercalation. In parallel, density functional theory (DFT) calculations are employed to evaluate adsorption geometries and energetics, providing a theoretical basis for interpreting the experimental data.

\section*{Results and Discussion}

\subsection*{N-heterocyclic carbenes on Au(111)}

The adsorption behavior of various NHCs on Au(111) has been extensively investigated experimentally under diverse conditions, including UHV and solution environments and even on different metallic substrates \cite{kaur2022fundamentals, trujillo2018using, crudden2014ultrastable, larrea2017, engel2017new, inayeh2021self, zhukhovitskiy2015carbene}. However, their role at the self-assembled monolayer (SAM) limit\cite{10.1002/chem.201903434, 10.1021/acsnano.0c01733} and in the context of intercalation remains unexplored. Here, we address this gap by providing both experimental and theoretical insight into the preferential surface interactions between NHC SAMs and GNRs. 
The N-substituent of the NHC significantly affects the adsorption geometry, with smaller groups favoring dimerization and bulkier ones promoting upright binding via adatoms \cite{inayeh2021self, kaur2022fundamentals}.
In this study, we investigate two NHC derivatives: a methyl-substituted NHC (NHC\textsuperscript{Me}), which is expected to form dimers,\cite{inayeh2021self} and an isopropyl-substituted NHC (NHC\textsuperscript{\textit{i}Pr}), which under room-temperature surface conditions is expected to adsorb as monomers in an upright geometry via single adatom coordination.\cite{bakker2018elucidating, inayeh2021self} Details of the synthesis and characterization of NHC\textsuperscript{Me} and NHC\textsuperscript{\textit{i}Pr} are provided in the Supporting Information (SI Section~\ref{sec:synthesis}).

\begin{figure}[H]
    \centering
    \includegraphics[width=\textwidth]{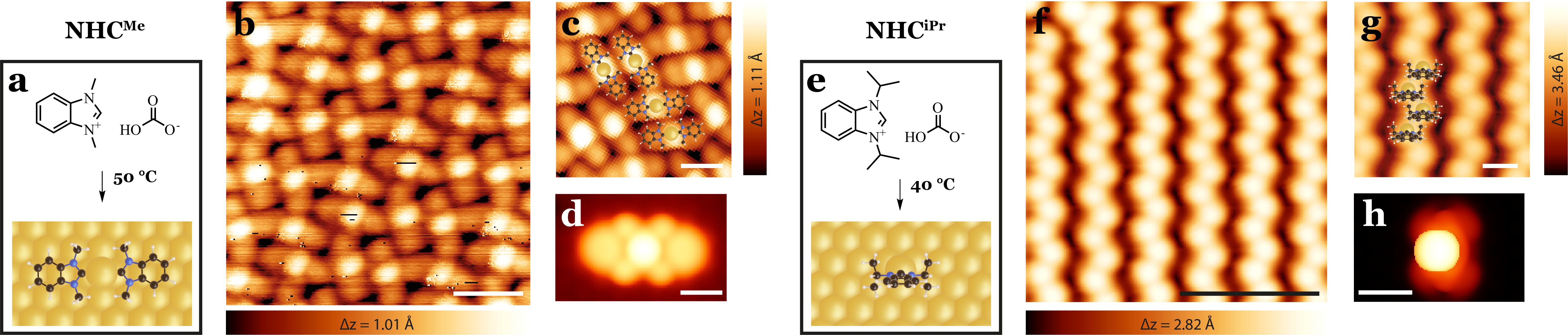}  
    \caption{%
    Structural and STM characterization of NHC\textsuperscript{Me} and NHC\textsuperscript{\textit{i}Pr} SAMs on Au(111). 
    \textbf{a} Molecular structure of NHC\textsuperscript{Me} and the dimer configuration formed via adatom coordination. 
    \textbf{b} Large-area STM image of the NHC\textsuperscript{Me} SAM, showing zigzagging dimer pairs. 
    \textbf{c} High-resolution STM image of NHC\textsuperscript{Me} dimers with an overlaid structural model. 
    \textbf{d} Simulated STM image of an isolated NHC\textsuperscript{Me} dimer on Au(111). 
    \textbf{e} Molecular structure of NHC\textsuperscript{\textit{i}Pr}, which binds upright to surface adatoms without forming dimers at room-temperature. 
    \textbf{f} STM image of the NHC\textsuperscript{\textit{i}Pr} SAM on Au(111), showing locally ordered zigzag rows of upright molecules. The packing is characterized by a rectangular unit cell with a staggered row arrangement.
    \textbf{g} High-resolution STM image of upright-standing NHC\textsuperscript{\textit{i}Pr} molecules with overlaid structure. 
    \textbf{h} Simulated STM image of the NHC\textsuperscript{\textit{i}Pr} SAM. 
    All STM images were acquired at 4~K. 
    STM scan parameters: 
    \textbf{b}: $V = -1$~V, $I = 40$~pA; 
    \textbf{c}: $V = -0.1$~V, $I = 50$~pA; 
    \textbf{f}, \textbf{g}: $V = -1$~V, $I = 30$~pA. 
    Scale bars: 
    \textbf{b}, \textbf{f}: 2~nm; 
    \textbf{c}, \textbf{g}: 1~nm;
    \textbf{d}, \textbf{h}: 0.5~nm.
    }
    \label{fig:NHC assemblies}
\end{figure}

The molecular structure of NHC\textsuperscript{Me} and its associated dimer configuration on Au(111) are shown in Figure~\ref{fig:NHC assemblies}\textbf{a}. The NHC\textsuperscript{Me} SAM was prepared in UHV by high-excess deposition at room temperature, using conditions that yield a complete monolayer (see Methods). 
Figure \ref{fig:NHC assemblies}\textbf{b} shows an STM image of the resulting NHC\textsuperscript{Me} SAM on Au(111), formed at room temperature.

NHC\textsuperscript{Me} molecules assemble into flat-lying dimer adsorbed structures on Au(111).
In STM, the dimers appear as recurring pairs in which two neighboring dimers are oriented approximately orthogonal ($\approx$90$^\circ$) to each other.
These orthogonal dimer pairs pack into quasi-parallel rows that are laterally offset with respect to adjacent rows, giving rise to the characteristic zigzag appearance of the monolayer (see SI Figure~\ref{fig:SI_unitcells}\textbf{a}).
The resulting assembly is described by an oblique unit cell of 3.18~nm~$\times$~3.09~nm with an internal angle of 63.9$^\circ$, containing eight flat-lying NHC\textsuperscript{Me} molecules. The SAM exhibits high local order. However, domain size varies significantly, forming along multiple directions, causing a lack of long-range order.
Figure \ref{fig:NHC assemblies}\textbf{c} shows a close-up of the dimers with the overlaid structural model. Figure \ref{fig:NHC assemblies}\textbf{d} depicts the STM simulation of an isolated dimer.

The NHC\textsuperscript{Me} SAM exhibits limited thermal stability, with desorption initiating below 100$~^\circ\mathrm{C}$. SI Figure~\ref{fig:SI_annealing_desorption} shows the evolution of the assembly with increasing temperature, reflecting the progressive loss of NHC\textsuperscript{Me} from the surface. Other reports on the desorption of comparable NHC SAMs suggest desorption occurring below 200$~^\circ\mathrm{C}$\cite{inayeh2021self, crudden2014ultrastable}. Molecules beyond monolayer coverage (weakly bound in an adlayer) desorb rapidly, already slightly above room-temperature. Consistent with this behavior, excess NHC\textsuperscript{Me} deposited during SAM formation gradually desorbs overnight at room temperature. Mild heating (below 50$~^\circ\mathrm{C}$) further accelerates this process, as slight annealing during deposition promotes rapid desorption of weakly bound molecules in the adlayer above the SAM. The remaining SAM is stable under UHV conditions, as confirmed by control measurements showing no change in the assembly after storing the sample for more than a week in UHV at room temperature (see SI Figure~\ref{fig:SI_RT_desorption}). 
This method is particularly advantageous for producing clean samples that are easier to scan and more suitable for further investigation on GNR intercalation.

To investigate the influence of the SAM adsorption geometry on the GNR intercalation yield, we also studied NHC\textsuperscript{\textit{i}Pr}, whose bulkier isopropyl group inhibits dimerization due to steric hindrance and instead leads to the formation of a SAM comprising upright-oriented molecules coordinated to surface adatoms.\cite{inayeh2021self, bakker2018elucidating, larrea2017, crudden2014ultrastable} The molecular structure of NHC\textsuperscript{\textit{i}Pr} is shown in Figure~\ref{fig:NHC assemblies}\textbf{e}. Figure \ref{fig:NHC assemblies}\textbf{f} shows the corresponding SAM of NHC\textsuperscript{\textit{i}Pr}, with upright features arranged in a quasi-hexagonal pattern, consistent with previous reports.\cite{bakker2018elucidating} Although surface contaminants disrupt long-range order, a local rectangular unit cell of approximately 1.7~nm~$\times$~0.69~nm (containing four upright NHC\textsuperscript{\textit{i}Pr} molecules) can be identified within clean regions. This rectangular unit cell arises from the offset zigzag arrangement of molecular rows, repeating every second row due to the staggered packing motif (see SI Figure~\ref{fig:SI_unitcells}\textbf{b}).
The acquired STM data for NHC\textsuperscript{\textit{i}Pr} also agrees well with the corresponding simulations, as shown in Figure~\ref{fig:NHC assemblies}\textbf{h}.

The discussed adsorption motifs are further supported by characteristic topographic features observed for both SAMs. STM images reveal surface vacancies in the Au(111) substrate (see SI Figure~\ref{fig:SI_holes}), which are attributed to the extraction of gold adatoms during carbene coordination.\cite{inayeh2021self, larrea2017} This adatom-mediated binding mechanism is consistent with previous reports and in excellent agreement with experimental STM data and corresponding simulations for both NHC\textsuperscript{Me} and NHC\textsuperscript{\textit{i}Pr} (Figures~\ref{fig:NHC assemblies}\textbf{d},\textbf{h}).

\subsection*{Assessing NHC Intercalation: Room- vs Low-Temperature Imaging}

\begin{figure}[H]
    \centering
    \includegraphics[width=\textwidth]{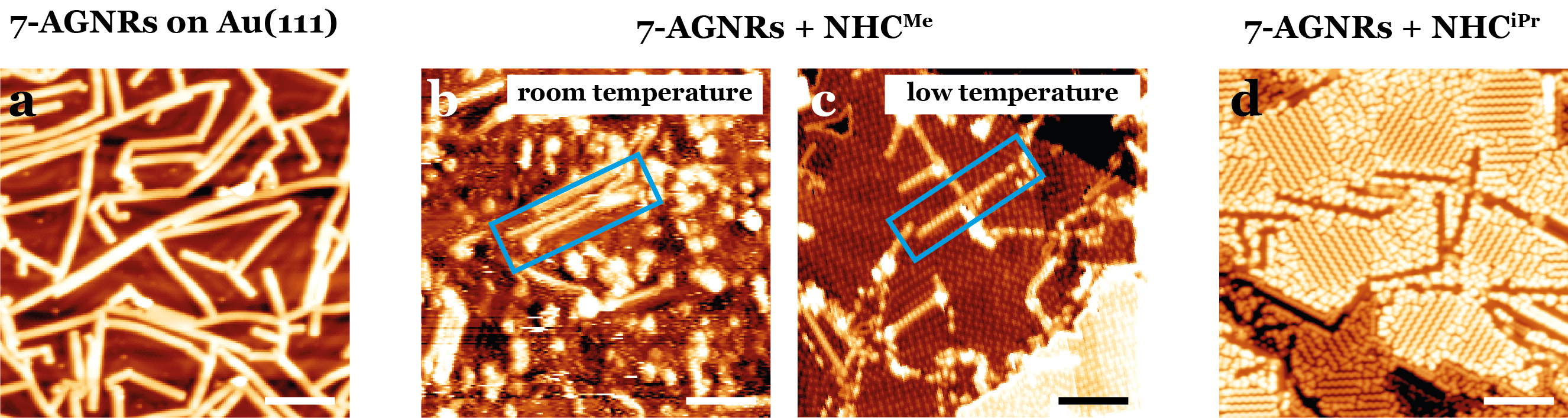}
    \caption{STM images related to attempts at intercalating 7-AGNRs with NHCs on Au(111). \textbf{a} 7-AGNRs grown on Au(111) by on-surface bottom-up synthesis under UHV conditions, imaged at 4~K. 
    \textbf{b} STM image acquired at room-temperature after deposition of NHC\textsuperscript{Me} onto a 7-AGNR-covered surface. Ribbon-like features (as highlighted in blue) are visible and appear to be located above the NHC\textsuperscript{Me} SAM, suggesting intercalation.
    \textbf{c} Similarly prepared sample imaged at 4~K. GNR-like features (as highlighted in blue) are again observed, though most appear segmented or discontinuous. 
    \textbf{d} STM image of 7-AGNRs with NHC\textsuperscript{\textit{i}Pr}, acquired at 4~K. The GNRs now appear more clearly embedded within the SAM, with a distinct height contrast between the two components.
    STM scan parameters:  \textbf{a-d}: $-1$~V, 20~pA; Scale bar: 10~nm.}
    \label{fig:NHCs GNR Intercalation}
\end{figure}

Prior to our intercalation experiments, we synthesized 7-AGNRs on Au(111) using a well-established bottom-up on-surface growth method (see Methods).\cite{cai2010atomically, talirz2017surface} As a prototypical system for our studies, we selected 7-AGNRs due to their well-characterized growth behavior, structural uniformity, and low chemical reactivity, which make them ideal candidates for proof-of-concept intercalation strategies. Figure~\ref{fig:NHCs GNR Intercalation}\textbf{a} shows a representative STM image of 7-AGNRs synthesized on Au(111), exhibiting medium surface coverage.

To prepare samples for intercalation, 7-AGNRs were grown on Au(111) and subsequently exposed to an excess of NHC\textsuperscript{Me}. After deposition, samples were left in UHV overnight to allow gradual desorption of excess NHC\textsuperscript{Me}, resulting in the formation of the stable SAM. Figure~\ref{fig:NHCs GNR Intercalation}\textbf{b} shows a room-temperature STM image of such sample. As expected, resolving the molecular features proved challenging, likely because of residual NHC mobility. However, GNR-like structures are discernible. At first glance, these features appear to lie atop the SAM, which would be consistent with successful intercalation.

This impression is supported by the apparent similarity in height (approximately 2\,\AA) between the SAM and the 7-AGNRs, as well as the uniform topographic appearance observed at room-temperature. Notably, the NHC\textsuperscript{Me} dimers and the 7-AGNRs exhibit comparable apparent heights in STM images, which makes it challenging to determine whether the measured height corresponds to the GNR resting on top of the SAM or to a derivative structure involving NHCs in a multilayer regime. Room-temperature STM can thus be ambiguous in the presence of mobile adsorbates, where weakly bound molecules or contaminants may interact with the surface and the tip, smoothing or masking local contrast.

To address this ambiguity, we repeated the experiment and acquired STM data at low temperature (4~K). Figure~\ref{fig:NHCs GNR Intercalation}\textbf{c} shows the resulting image. The overall morphology remains similar to that observed at room-temperature, with GNR-like structures still discernible atop the SAM. However, suppressing thermal diffusion reveals additional structural details: many apparent ribbons now display a segmented topography absent from the room-temperature images. This indicates that the smoother appearance observed at room-temperature arises from dynamic averaging due to the mobility of weakly bound surface species.

The low-temperature data, therefore, resolve a key ambiguity in the room-temperature interpretation. Apparent intercalated 7-AGNRs likely originate from another adsorbed species rather than genuine intercalated ribbons. These results highlight an important experimental caveat: STM measurements at room-temperature can lead to misinterpretations of molecular arrangement in systems with mobile adsorbates. Complementary low-temperature imaging is thus essential to accurately resolve the true surface structure in such cases.

In the case of NHC\textsuperscript{\textit{i}Pr}, only low-temperature STM measurements were performed. Figure~\ref{fig:NHCs GNR Intercalation}\textbf{d} shows a representative image of a sample prepared by depositing NHC\textsuperscript{\textit{i}Pr} onto previously grown 7-AGNRs on Au(111). In contrast to the NHC\textsuperscript{Me} case, the GNRs now appear fully embedded within the surrounding SAM, with a distinct height difference that clearly separates the ribbons from the NHC\textsuperscript{\textit{i}Pr} monolayer. No GNR-like features are observed that extend above the surface, suggesting that NHC\textsuperscript{\textit{i}Pr} molecules do not adsorb atop the ribbons. This points to fundamental energetic or steric differences between the two NHCs that may suppress intercalation for NHC\textsuperscript{\textit{i}Pr}.

However, the STM data also reveal the presence of surface impurities or disorder, likely related to the low deposition temperatures required for NHC\textsuperscript{\textit{i}Pr}. As discussed earlier, this contamination limits large-scale SAM ordering,\cite{bakker2018elucidating} which may in turn reduce the efficiency or uniformity of intercalation. Although local embedding of GNRs is clearly observed, the disrupted assembly prevents us from drawing definitive conclusions about the overall intercalation yield for the NHC\textsuperscript{\textit{i}Pr} case.

\subsection*{7-AGNRs Intercalated with NHC\textsuperscript{Me}}

To clarify the intercalation behavior of 7-AGNRs in the NHC\textsuperscript{Me} system, where adsorbed species may obscure the actual appearance of the structural configuration, we performed high-resolution low-temperature STM measurements. These images revealed three distinct configurations of the 7-AGNRs and the NHC\textsuperscript{Me} layer: (i) 7-AGNRs with adsorbed NHC\textsuperscript{Me} dimers on top, (ii) 7-AGNRs embedded within the SAM without additional species, and (iii) 7-AGNRs that appear fully intercalated above the NHC\textsuperscript{Me} layer.

\begin{figure}[H]
    \centering
    \includegraphics[width=0.6\textwidth]{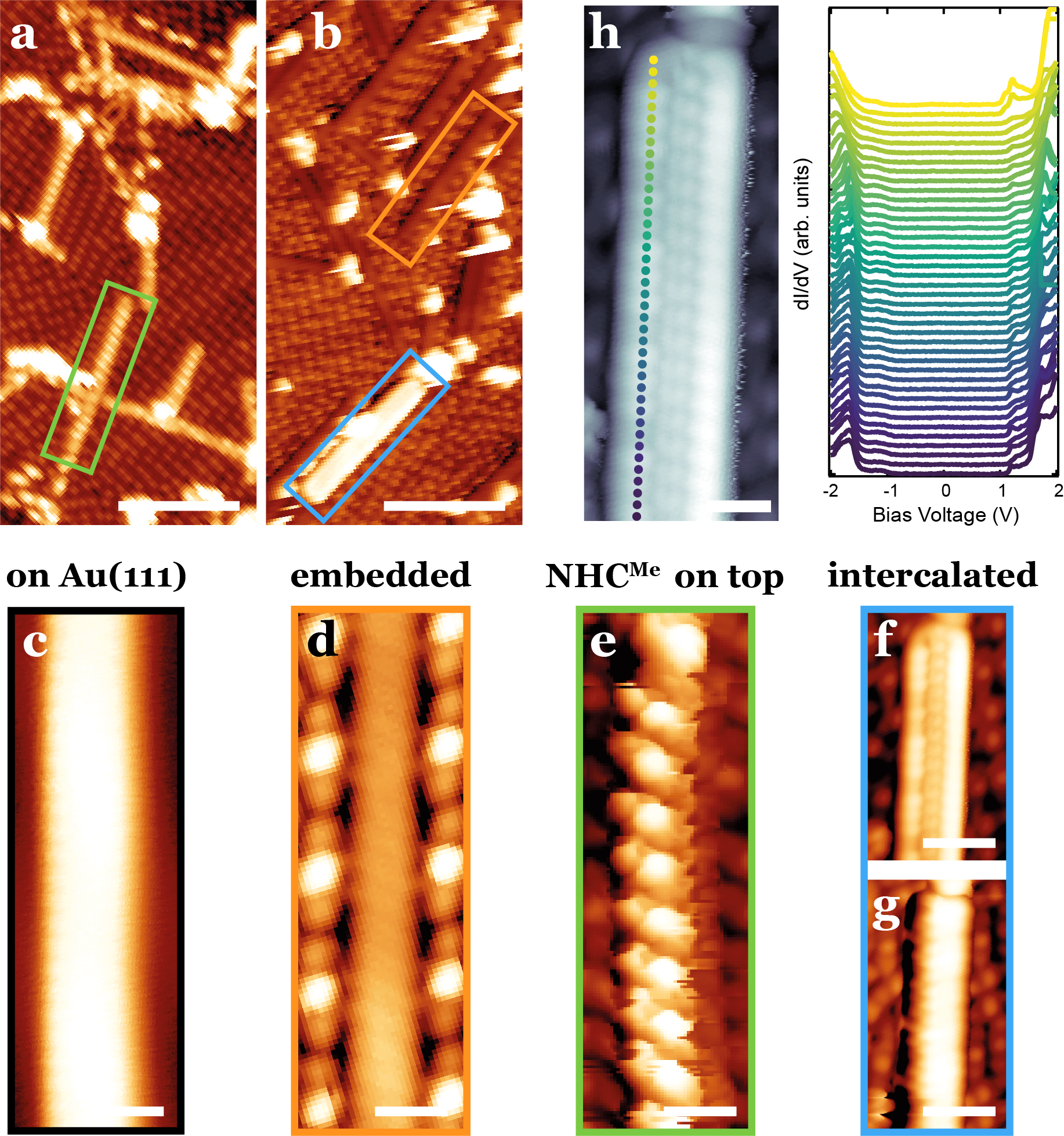}  
    \caption{Low-temperature STM analysis of 7-AGNRs combined with NHC\textsuperscript{Me} SAMs on Au(111), revealing multiple interaction motifs. 
    \textbf{a} GNR with adsorbed NHC\textsuperscript{Me} dimers on its surface, evident from the segmented, dimer-like contrast along the ribbon axis. 
    \textbf{b} Sample after mild annealing (below 50$~^\circ$C), showing cleaner embedded GNRs free of adsorbates. 
    \textbf{c} Reference image of a 7-AGNR on bare Au(111). 
    \textbf{d} Close-up showing densely packed NHC\textsuperscript{Me} dimers adsorbed along the GNR edges. 
    \textbf{e} Embedded GNR decorated with densely packed adsorbates on top.
    \textbf{f} Intercalated GNR at high negative bias, showing the typical valence band contrast of electronically decoupled 7-AGNRs. 
    \textbf{g} Same GNR imaged at lower bias, where the state vanishes, confirming its voltage dependence. 
    \textbf{h} d$I$/d$V$ spectrum recorded on an intercalated GNR, showing a symmetric band gap centered at 0~V, consistent with decoupling from the substrate. STM scan parameters: \textbf{a}, \textbf{b}, \textbf{f}: $-2$~V, 20~pA; 
    \textbf{c}: $-1$~V, 1~nA; 
    \textbf{d}: $-0.1$~V, 50~pA; 
    \textbf{e}: $-0.1$~V, 20~pA; 
    \textbf{g}: $-0.5$~V, 20~pA. 
    Scale bars: \textbf{a}, \textbf{b}: 10~nm; \textbf{c}–\textbf{h}: 1~nm.}
    \label{fig:7-AGNRs + NHCMe Intercalation in LT}
\end{figure}

Initially, many 7-AGNRs appeared to exhibit the aforementioned segmentation along their axis (Figure~\ref{fig:7-AGNRs + NHCMe Intercalation in LT}\textbf{a},\textbf{e}). Upon close inspection, this apparent substructure matched precisely with the known shape of NHC\textsuperscript{Me} dimers in the SAM (Figure \ref{fig:NHC assemblies}\textbf{a-c}), indicating that excess NHC\textsuperscript{Me} were adsorbed on top of the 7-AGNRs. These dimers preferentially orient perpendicular to the long axis of the GNR, stacking densely due to their close width match with the underlying 7-AGNR. As a result, they mimic the apparent geometry of the ribbon. Importantly, these 7-AGNRs are not intercalated but rather embedded within the SAM, with additional NHC\textsuperscript{Me} dimers physisorbed on top. This adsorption configuration complicates the identification of truly intercalated structures. This suggests that GNRs act as preferential adsorption sites for excess NHC\textsuperscript{Me}, possibly stabilizing otherwise weakly bound dimers and inhibiting their desorption.

To selectively remove weakly bound adsorbates and test the stability of the remaining structures, we annealed the sample at mild temperatures (below 50$^\circ$C), which desorbs the adlayer as previously established.\cite{inayeh2021self} This treatment revealed clean GNR segments embedded within the SAM, free of additional adsorbates (Figure~\ref{fig:7-AGNRs + NHCMe Intercalation in LT}\textbf{b},\textbf{d}). Notably, a subset of GNR-like features remained after annealing. These appeared topographically distinct, lacking the segmentation characteristic of 7-AGNRs decorated with adsorbed dimers (Figure~\ref{fig:7-AGNRs + NHCMe Intercalation in LT}\textbf{f},\textbf{g}). We attribute these structures to truly intercalated 7-AGNRs above the NHC\textsuperscript{Me} SAM, effectively decoupled from the Au(111) substrate. This interpretation is supported by STM imaging at high bias (Figure~\ref{fig:7-AGNRs + NHCMe Intercalation in LT}\textbf{f}), where we observe a strong occupied state at $-2$~V, typically suppressed when ribbons are directly adsorbed on Au(111). At lower bias of $-0.5$~V (Figure \ref{fig:7-AGNRs + NHCMe Intercalation in LT}\textbf{g}), this feature vanishes.

Further evidence for intercalation comes from scanning tunneling spectroscopy (STS) acquired on the intercalated GNR (Figure~\ref{fig:7-AGNRs + NHCMe Intercalation in LT}\textbf{h}). The dI/dV spectra reveal a symmetric band gap of  2.8 eV centered at 0 V, in contrast to ribbons on Au(111) with a gap of 2.4 eV where Fermi-level pinning often leads to asymmetric HOMO/LUMO alignment and broadening of the frontier resonances.\cite{talirz2017surface, 10.1002/cphc.201900313} A related example is provided by Si-intercalated Au(111), where the formation of a gold silicide buffer suppresses the Au surface state and enables clearer resolution of the frontier states, though still with residual asymmetry in level alignment.\cite{deniz2017revealing, ruffieux2012electronic} In our case, the observed symmetry and absence of charging features suggest that the intercalated 7-AGNRs are effectively decoupled from the metallic substrate by the NHC\textsuperscript{Me} layer. This behaviour is closely analogous to that reported for 9-AGNRs transferred onto epitaxial graphene, which likewise exhibit symmetric valence and conduction band alignment with respect to the Fermi level and reduced substrate screening.\cite{doi:10.1021/acsanm.5c02753} The resulting symmetry indicates that the ribbons remain uncharged upon intercalation, highlighting the suitability of NHC-based decoupling for preserving the intrinsic electronic structure of 7-AGNRs.

While these findings confirm the feasibility of NHC-assisted GNR intercalation, the process appears to be inefficient under the current conditions. Based on the total length of GNR segments that are embedded or intercalated, the yield of intercalation is estimated to be only around 1.35\% (see SI Figure~\ref{fig:SI_Yield_Analysis}). This low yield suggests that intercalation is highly sensitive to preparation conditions and may require further optimization to increase efficiency.

\subsection*{Simulations of different NHC/GNR Configurations}
 
\begin{figure*}[h!]
\centering
\includegraphics[width=0.9\textwidth]{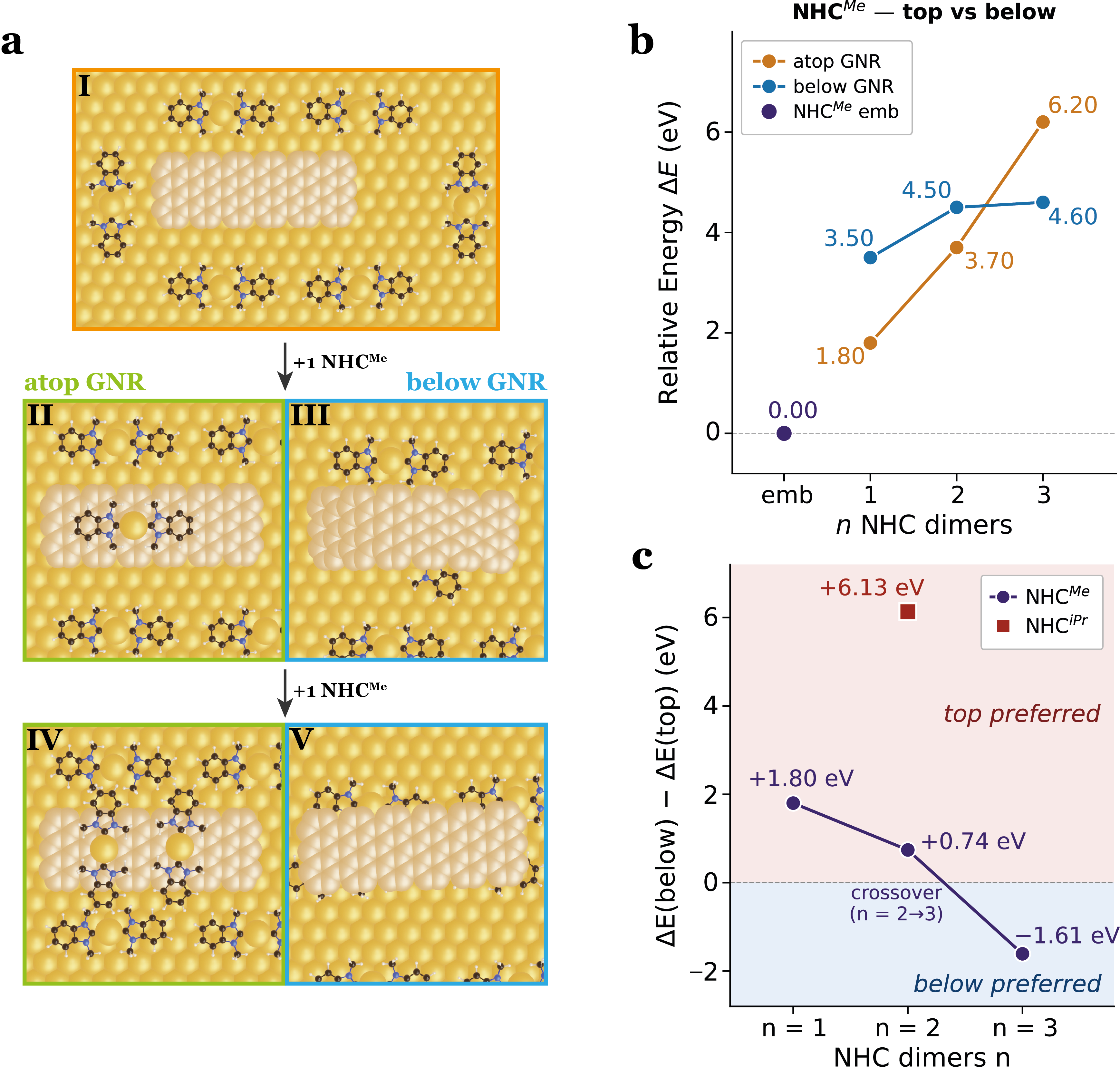}
\caption{%
\textbf{a} DFT-optimized geometries of different NHC\textsuperscript{Me}--7-AGNR configurations.
\textbf{I} Embedded 7-AGNR in a SAM of six NHC\textsuperscript{Me} dimers on Au(111), serving as the reference configuration.
\textbf{II--V} Subsequent configurations illustrate adsorption and intercalation steps as discussed in the main text.
\textbf{b} Relative energies ($\Delta E$) of relaxed configurations for NHC\textsuperscript{Me}, referenced to the embedded structure (emb). ``atop
GNR'' denotes configurations with one, two, or three NHC units adsorbed on top of the ribbon; ``below GNR'' denotes configurations with one, two, or three NHC
units placed below the GNR.
\textbf{c} Difference in relative energy between the below-ribbon and atop-ribbon configurations, $\Delta E(\text{below}) - \Delta E(\text{top})$, as
a function of the number of displaced NHC\textsuperscript{Me} units $n$. Positive values indicate that the atop configuration is preferred; negative values indicate
that intercalation (below) becomes energetically favorable. Configurations that do not relax to stable target geometries are not included.}
\label{fig:simulations}
\end{figure*}

To rationalize the experimental observations and evaluate the relative energetic accessibility of different adsorption and intercalation motifs, we performed DFT-based geometry optimizations for a six-anthracene-unit-long 7-AGNR on Au(111) in the presence of NHCs. Separate calculations were carried out for NHC\textsuperscript{Me} and NHC\textsuperscript{\textit{i}Pr}. In each case, the embedded configuration within the NHC-SAM was used as the reference structure ($\Delta E = 0$~eV). The reported $\Delta E$ values therefore describe relative energies of relaxed metastable configurations and do not represent kinetic activation barriers. Figure~\ref{fig:simulations}\textbf{a} shows the
optimized configurations for NHC\textsuperscript{Me}, with panel~\textbf{I} depicting the embedded reference geometry. The corresponding optimized structures for NHC\textsuperscript{\textit{i}Pr} are provided in the Supporting Information (SI Figure~\ref{fig:SI_NHCiPr_simulations}).
 
Starting from the embedded reference structure, we considered stepwise rearrangements in which NHC units were displaced either onto the GNR or
underneath it, i.e.\ between the ribbon and the Au(111) substrate. Figure~\ref{fig:simulations}\textbf{b} shows the resulting relative energies for
NHC\textsuperscript{Me}. For adsorption on top of the ribbon, the relative energy increases approximately linearly with the number of displaced NHC units. The
relaxed top-adsorbed structures lie at $\Delta E = 1.80$, 3.70, and 6.20~eV for one, two, and three units on top, respectively. Thus, adsorption on top of the
ribbon remains an accessible metastable motif but becomes progressively less favorable as more NHCs are forced into the confined space above the GNR.
 
A different trend is found when NHC\textsuperscript{Me} units are inserted below the ribbon. Placing the first unit underneath the GNR raises the energy to $\Delta
E = 3.50$~eV, indicating that the initial intercalation step carries the main energetic penalty, consistent with the need to locally lift and distort the
otherwise planar GNR and to reduce its van der Waals contact with the Au(111) surface. In contrast, insertion of additional NHC\textsuperscript{Me} units below
the already-lifted ribbon causes only a small further increase in energy, with the two-below and three-below structures found at $\Delta E = 4.50$ and 4.60~eV, respectively. Once the initial lifting penalty has been paid, further intercalation becomes markedly less costly.
 
Figure~\ref{fig:simulations}\textbf{c} directly compares the below and atop configurations by plotting $\Delta E(\text{below}) - \Delta E(\text{top})$ as a
function of $n$. For $n = 1$, the below configuration is substantially less favorable than the top-adsorbed state ($+1.80$~eV). For $n = 2$, this gap
decreases considerably ($+0.74$~eV). For $n = 3$, the trend reverses: the below configuration becomes lower in energy than the top-adsorbed one ($-1.61$~eV),
indicating a crossover between $n = 2$ and $n = 3$. Within the present model, this shows that once the ribbon has been locally lifted, continued accommodation
of additional NHC\textsuperscript{Me} units beneath the GNR becomes increasingly competitive with, and eventually more favorable than, further crowding above the ribbon.
 
For NHC\textsuperscript{\textit{i}Pr}, the situation is markedly different. The top-adsorbed configurations follow a similar trend to NHC\textsuperscript{Me},
with relaxed structures at $\Delta E = 1.96$, 4.02, and 6.26~eV for one, two, and three units on top, respectively. The below-ribbon configurations, however, are far less accessible. When a single NHC\textsuperscript{\textit{i}Pr} unit is placed below the GNR, the structure does not relax to a stable intercalated configuration but reverts
to the embedded geometry. Attempts to generate a three-below configuration likewise produced no stable minimum, with the system again relaxing back to the
embedded structure. Because no distinct metastable below-ribbon state exists for $n = 1$ or $n = 3$, no meaningful $\Delta E(\text{below}) - \Delta E(\text{top})$
difference can be defined for those cases, and panel~\textbf{c} therefore contains only a single NHC\textsuperscript{\textit{i}Pr} data point. For $n = 2$, a
below-ribbon configuration can be stabilized, but only at $\Delta E = 7.18$~eV, substantially above the corresponding top-adsorbed structure at 4.02~eV. The
resulting difference of $+6.13$~eV lies well within the top-preferred regime, as shown in Figure~\ref{fig:simulations}\textbf{c}. These results show that intercalation of NHC\textsuperscript{\textit{i}Pr} is significantly less favorable than for NHC\textsuperscript{Me}.

\subsection*{Raman Spectroscopy of 7-AGNRs with NHC SAMs}

Raman spectroscopy provides access to the characteristic vibrational fingerprints of GNRs. In particular, the G mode ($\sim$1600~cm$^{-1}$) arises from in-plane C–C stretching, while the D mode ($\sim$1340~cm$^{-1}$) originates from edge-induced breathing vibrations of the sp$^2$ lattice and is intrinsic to nanoribbons rather than defect-activated as in graphene.\cite{casiraghi2017raman} Additional edge-localized C–H bending modes appear between 1200–1300~cm$^{-1}$, and the low-frequency radial breathing-like mode (RBLM) directly reflects the ribbon width. Together, these features provide a sensitive probe of ribbon structure, edge integrity, and ultimately, in our case, the interaction with the NHC environment.\cite{verzhbitskiy2016raman, borin2019surface, barin2023surface}

\begin{figure}[H]
    \centering
    \includegraphics[width=\textwidth]{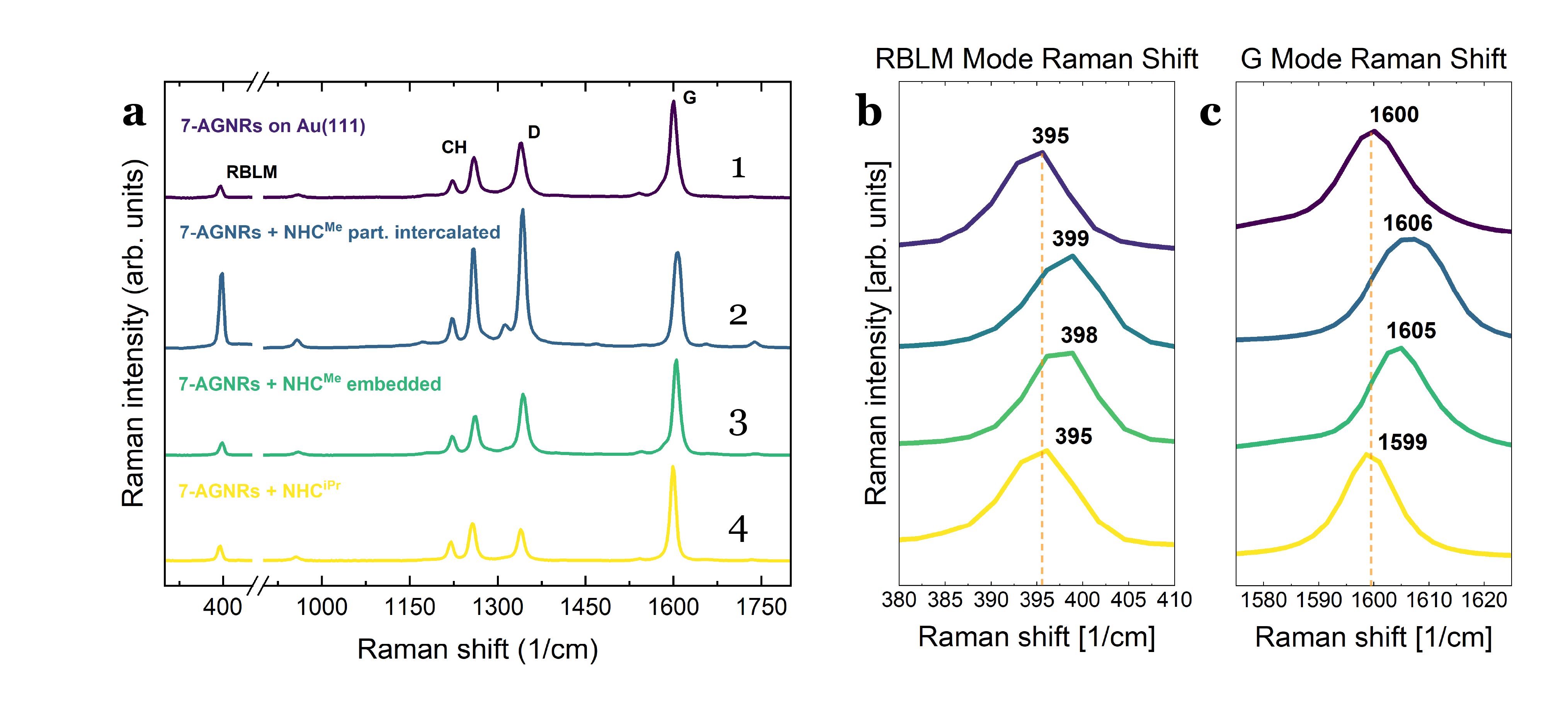}  
    \caption{Raman spectroscopic characterization of 7-AGNRs with and without NHC intercalation or embedding. 
    \textbf{a} Raman spectra acquired using a 532~nm laser (spectral range: 320–1790~cm\textsuperscript{-1}) for: 
    (1) 7-AGNRs on Au(111) (laser power: 10~mW, integration time: 0.5~s), 
    (2) 7-AGNRs with excess NHC\textsuperscript{Me} resulting in partial intercalation (10~mW, 0.5~s), 
    (3) 7-AGNRs embedded in an NHC\textsuperscript{Me} SAM following mild annealing to desorb excess carbenes (20~mW, 0.5~s), and 
    (4) 7-AGNRs embedded in an NHC\textsuperscript{\textit{i}Pr} SAM (2~mW, 1~s). 
    \textbf{b} Close-up of the RBLM region. A shift to higher wavenumbers is observed for samples with NHC\textsuperscript{Me} (2 and 3), while no shift is detected for the NHC\textsuperscript{\textit{i}Pr} case (4). 
    \textbf{c} Close-up of the G-mode region. Similar shifts are observed as in the RBLM.
    }
    \label{fig:Raman spectroscopy}
\end{figure}

Figure~\ref{fig:Raman spectroscopy}\textbf{a} presents spectra from four representative sample configurations, recorded under UHV conditions using a 532~nm laser. As a reference, 7-AGNRs directly on Au(111) (spectrum~1) exhibit the expected dominant G mode, along with clearly resolved C--H, D, and RBLM features. When 7-AGNRs are exposed to an excess of NHC\textsuperscript{Me} (spectrum~2), the spectral profile changes markedly: the relative intensities of the C--H, D, and RBLM modes increase compared to the G mode, and the main Raman features shift to \emph{higher} wavenumbers (Figure~\ref{fig:Raman spectroscopy}\textbf{b--c}). Quantitatively, the RBLM maximum shifts from $\sim$395~cm$^{-1}$ (spectrum~1) to $\sim$399~cm$^{-1}$ (spectrum~2), while the G mode shifts from $\sim$1600~cm$^{-1}$ to $\sim$1606~cm$^{-1}$. Such shifts can originate from changes in charge transfer, screening, strain, or substrate coupling, and have been widely reported in graphene-based systems under doping and modified interfacial conditions.\cite{das2008monitoring, das2008raman, eckmann2012probing, wang2016giant}

To disentangle electronic effects from structural changes due to intercalation, a third sample was prepared in which mild annealing removed excess NHC\textsuperscript{Me}, leaving 7-AGNRs embedded within the SAM (spectrum~3). Here, the peak positions remain shifted relative to the bare Au(111) reference, but slightly less than in spectrum~2: the RBLM is at $\sim$398~cm$^{-1}$ and the G mode at $\sim$1605~cm$^{-1}$ (Figure~\ref{fig:Raman spectroscopy}\textbf{b--c}). In contrast to spectrum~2, however, the relative mode intensities largely revert towards those of the reference, suggesting that the pronounced intensity changes in spectrum~2 are not simply a consequence of a global electronic shift, but correlate with the partially intercalated configuration. This is consistent with earlier reports and calculations showing that the relative intensities of GNR Raman modes can be particularly sensitive to substrate interactions, screening, and coupling conditions.\cite{overbeck2019universal} At the same time, the persistence of the frequency shift in spectrum~3 indicates that electronic effects, such as residual charge transfer or altered screening, may still contribute.\cite{hell2018resonance,basko2009electron}

Further insight is obtained from the fourth sample, where 7-AGNRs are embedded in an NHC\textsuperscript{\textit{i}Pr} SAM (spectrum~4). As discussed above, NHC\textsuperscript{\textit{i}Pr} does not intercalate the nanoribbons. In this case, the Raman spectra do not exhibit a comparable upshift of the main vibrational modes. The RBLM remains at approximately 395~cm$^{-1}$ and the G mode at about 1599~cm$^{-1}$, which is close to the position observed for 7-AGNRs directly on Au(111) within the experimental uncertainty. This comparison is relevant because NHC layers are known to modify the gold work function through the formation of an interface dipole,\cite{kim2017reduction} which could in principle influence charge transfer and thereby affect Raman peak positions. If the frequency shifts observed for NHC\textsuperscript{Me} were governed predominantly by a work-function-driven doping effect, a similar trend might also be expected for NHC\textsuperscript{\textit{i}Pr}. The absence of such a shift suggests that the upshift observed for NHC\textsuperscript{Me} in spectrum~2 and spectrum~3 is more closely related to the intercalation and the associated change in substrate coupling and screening. At the same time, we cannot exclude that additional factors such as local strain, residual adsorbates, or variations in electronic screening contribute to the measured peak positions.

Consistent with the spectral trends, analysis of mode intensity ratios further supports the link to partial intercalation. Only the partially intercalated sample (spectrum~2) exhibits C--H/G and D/G ratios exceeding unity (see SI Figure~\ref{fig:SI_iPr_Temp_Raman}), whereas all other cases—7-AGNRs on Au(111), 7-AGNRs embedded in an NHC\textsuperscript{Me} SAM after annealing (spectrum~3), and 7-AGNRs embedded in an NHC\textsuperscript{\textit{i}Pr} SAM (spectrum~4)—remain below one and follow a consistent trend. Taken together, the combination of (i) increased relative C--H/D/RBLM intensities and (ii) an upshift of the RBLM and G mode provides a practical spectroscopic fingerprint of the partially intercalated configuration in our samples.

While the observed spectral changes are consistent with intercalation and decoupling, they should not be viewed as conclusive on their own. Other factors such as local strain, adsorbates, or variations in electronic screening could also influence the Raman response. Still, the strong correlation between these spectroscopic features and independently observed structural differences suggests that they may serve as valuable indicators of intercalation. This approach could offer a practical, non-invasive means to assess interfacial structure and coupling in graphene-based systems.

\section*{Conclusion}

We have shown that NHC\textsuperscript{Me} SAMs can partially intercalate 7-AGNRs on Au(111), producing locally decoupled segments at low yield. We identified distinct adsorption configurations and elucidated how the NHC geometry affects their assembly. In particular, NHC\textsuperscript{Me} facilitates partial intercalation, leading to ribbons that are electronically decoupled from the gold substrate. In contrast, the bulkier NHC\textsuperscript{\textit{i}Pr} forms upright SAMs that appear to prevent intercalation under comparable conditions.

Importantly, the low intercalation yield in the NHC\textsuperscript{Me} system suggests that the process operates at a threshold regime, where small changes in molecular geometry, binding energetics, or local packing may critically influence the intercalation outcome. This intermediate regime presents a valuable opportunity to study the chemical and structural parameters that govern intercalation performance. The combination of experimental tools with theoretical simulations used here provides a robust framework for probing these factors in detail.

Looking forward, our findings establish a basis for the rational design of more efficient intercalants. By tailoring ligand bulk, binding strength, and mobility—guided by insights from systems like NHCs—future studies may enhance intercalation yield and uniformity. Such advances would open new pathways toward scalable decoupling and dry-transfer under UHV conditions of GNRs and related nanostructures.

\section*{Methods}

\subsection*{Organic Synthesis - General Methods and Materials}

Detailed experimental procedures for the synthesis of NHC\textsuperscript{Me} and NHC\textsuperscript{\textit{i}Pr}, including characterization data, are provided in SI Section \ref{sec:synthesis}.

All reagents were obtained from Sigma Aldrich, TCI, abcr, Alfa Aesar, and Strem, and were used as received without further purification.

NMR spectra were recorded on a Bruker AV-II 300 spectrometer operating at 300 MHz for $^1$H and 75 MHz for $^{13}$C. D$_2$O ($\delta(^1$H) = 4.79 ppm) was used as solvent and as the internal chemical shift reference. Chemical shifts ($\delta$) are reported in ppm. The following abbreviations are used to describe peak multiplicities: s = singlet, d = doublet, t = triplet, q = quartet, and m = multiplet.

\subsection*{Sample Preparation}

Gold single crystals (Au(111)) were cleaned via repeated sputtering (Ar$^+$, 1 keV, 10 min) and annealing cycles (450$\,^{\circ}$C, 10 min) in UHV. For the synthesis of 7-AGNRs, molecular precursors (10,10'-dibromo-9,9'-bianthryl, DBBA)\cite{cai2010atomically} were deposited using a Knudsen cell at 170$\,^{\circ}$C. Samples were annealed to 200$\,^{\circ}$C to induce polymerization, followed by a second annealing step at 400$\,^{\circ}$C for cyclodehydrogenation.

NHC derivatives were generated in situ by thermal decomposition of their bicarbonate adduct precursors and deposited onto the surface from a dedicated Knudsen cell under UHV conditions. For NHC\textsuperscript{Me}, deposition was performed at room temperature with the crucible heated until the chamber pressure reached $1\times10^{-9}$ mbar. For NHC\textsuperscript{\textit{i}Pr}, lower crucible temperatures were used due to its higher volatility. In all cases, samples were left in UHV overnight to allow spontaneous desorption of excess physisorbed molecules. For faster preparation, annealing below 50$\,^{\circ}$C was used to promote clean SAM formation without decomposition.

\subsection*{STM Imaging}

STM measurements were performed using an Omicron VT-STM operated at room temperature and Omicron LT-STM at 4 K, both operated under UHV conditions. Electrochemically etched tungsten tips were used. Images were acquired in constant current mode. All scan parameters and scale bars are included in the figures.

\subsection*{Raman Spectroscopy}

Raman spectroscopy measurements were performed using a WITec Alpha~300~R confocal Raman microscope operated in backscattering geometry.
A 50$\times$ long working distance objective (numerical aperture NA~=~0.55, working distance~=~9.1~mm) was used to focus the laser onto the sample and collect the scattered light.
Excitation was provided by a 532~nm laser.

The backscattered Raman signal was detected without an analyzing polarizer and coupled to a lens-based spectrometer equipped with a cooled CCD detector.
Laser power and integration time were optimized for each sample to maximize signal intensity while minimizing laser-induced damage.
All measurement parameters, including laser power and integration time, are specified in the corresponding figure captions. A polynomial background was subtracted from the raw spectra.

Measurements were carried out under vacuum using an in-house built portable Raman vacuum suitcase, pumped by an ion pump and maintaining pressures in the range of high $10^{-8}$ to low $10^{-7}$~mbar.
The vacuum chamber was mounted on a motorized stage to allow spatial positioning of the sample during measurement.

\subsection*{Simulations}

All calculations were performed with AiiDAlab~\cite{aiidalab} apps based on AiiDA~\cite{aiida} workflows for the DFT code CP2K~\cite{cp2k}. The surface/adsorbate system was modeled within the repeated slab scheme. The simulation cell contained 2322 atoms for all NHC\textsuperscript{\textit{i}Pr} configurations and 2172 for all NHC\textsuperscript{Me} configurations, including four atomic layers of Au along the [111] direction and a layer of hydrogen atoms to passivate one side of the slab in order to suppress one of the two Au(111) surface states. 40 Å of vacuum was included in the simulation cell to decouple the system from its periodic replicas in the direction perpendicular to the surface. The size of the cell was 59.0×46.0 Å$^2$ corresponding to 180 Au(111) surface unit cells. The electronic states were expanded with a TZV2P Gaussian basis set~\cite{basis} for C and H species and a DZVP basis set for Au. A cutoff of 600 Ry was used for the plane wave basis set. Norm-conserving Goedecker-Teter-Hutter pseudopotentials~\cite{willand2013norm} were used to represent the frozen core electrons of the atoms. We used the PBE parameterization for the generalized gradient approximation of the exchange correlation functional~\cite{pbe}. To account for van der Waals interactions, we used the D3 scheme proposed by Grimme~\cite{grimme2010consistent, vdwd3}. To obtain the equilibrium geometries, we kept the atomic positions of the bottom two layers of the slab fixed to the ideal bulk positions, and all other atoms were relaxed until forces were lower than 0.005 eV/Å.
STM images were simulated within the Tersoff-Hamann approximation \cite{Tersoff1993} based on the Kohn–Sham orbitals of the slab/adsorbate systems. The orbitals were extrapolated to the vacuum region in order to correct the wrong decay of the charge density due to the localized basis set.

\subsection*{Yield Estimation}

Intercalation yield was estimated by manual annotation of STM images. For each sample, the total length of visible ribbons was measured, and the fraction exhibiting intercalated appearance was calculated. Only high-resolution images acquired after annealing at 50$\,^{\circ}$C were considered. This approach gives a lower-bound estimate of yield (see SI Figure~\ref{fig:SI_Yield_Analysis}).

\section*{Acknowledgements}

Dominik L\"uthi, Gabriela Borin Barin, and Roman Fasel acknowledge financial support from the Werner Siemens Foundation (CarboQuant project). Gabriela Borin Barin and Roman Fasel further acknowledge funding from the European Union’s Horizon Europe research and innovation program under Grant Agreement No. 101099098 (ATYPIQUAL), as well as from the State Secretariat for Education, Research and Innovation (SERI) under Contract No. 23.00422. 

Gabriela Borin Barin acknowledges support from the Swiss National Science Foundation (SNSF) under Grant No. 200021E-219172/1 (GRAAL). Dominik L\"uthi acknowledges financial support from the University of Bern.

\bibliography{main}

@article{10.1021/acs.jpcc.5b00528,
  author  = {Fei, Xiangmin and Zhang, Lizhi and Xiao, Wende and Chen, Hui and Que, Yande and Liu, Liwei and Yang, Kai and Du, Shixuan and Gao, Hong-Jun},
  title   = {Structural and Electronic Properties of Pb-Intercalated Graphene on {Ru(0001)}},
  journal = {The Journal of Physical Chemistry C},
  year    = {2015},
  doi     = {10.1021/acs.jpcc.5b00528},
}

@article{10.1088/1674-1056/ac6941,
  author  = {Peng, Hao and Jin, Xin and Song, Yang and Du, Shixuan},
  title   = {First Principles Study of Hafnium Intercalation Between Graphene and {Ir(111)} Substrate},
  journal = {Chinese Physics B},
  year    = {2022},
  doi     = {10.1088/1674-1056/ac6941},
}

@article{el2020controlled,
  title={Controlled quantum dot formation in atomically engineered graphene nanoribbon field-effect transistors},
  author={El Abbassi, Maria and Perrin, Mickael L and Barin, Gabriela Borin and Sangtarash, Sara and Overbeck, Jan and Braun, Oliver and Lambert, Colin J and Sun, Qiang and Prechtl, Thorsten and Narita, Akimitsu and others},
  journal={ACS nano},
  volume={14},
  number={5},
  pages={5754--5762},
  year={2020},
  publisher={ACS Publications}
}

@article{10.1002/chem.201903434,
  author  = {Dery, Shahar and Kim, Suhong and Tomaschun, Gabriele and Haddad, David and Cossaro, Albano and Verdini, Alberto and Floreano, Luca and Kl{\"u}ner, Thorsten and Toste, F. Dean and Gross, Elad},
  title   = {Flexible {NO}$_2$-Functionalized {N}-Heterocyclic Carbene Monolayers on {Au(111)} Surface},
  journal = {Chemistry -- A European Journal},
  year    = {2019},
  doi     = {10.1002/chem.201903434},
}

@article{yamaguchi2020small,
  title={Small bandgap in atomically precise 17-atom-wide armchair-edged graphene nanoribbons},
  author={Yamaguchi, Junichi and Hayashi, Hironobu and Jippo, Hideyuki and Shiotari, Akitoshi and Ohtomo, Manabu and Sakakura, Mitsuhiro and Hieda, Nao and Aratani, Naoki and Ohfuchi, Mari and Sugimoto, Yoshiaki and others},
  journal={Communications Materials},
  volume={1},
  number={1},
  pages={36},
  year={2020},
  publisher={Nature Publishing Group UK London}
}

@article{xiang2025zigzag,
  title={Zigzag graphene nanoribbons with periodic porphyrin edge extensions},
  author={Xiang, Feifei and Gu, Yanwei and Kinikar, Amogh and Bassi, Nicol{\`o} and Ortega-Guerrero, Andres and Qiu, Zijie and Gr{\"o}ning, Oliver and Ruffieux, Pascal and Pignedoli, Carlo A and M{\"u}llen, Klaus and others},
  journal={Nature Chemistry},
  volume={17},
  number={9},
  pages={1356--1363},
  year={2025},
  publisher={Nature Publishing Group UK London}
}

@article{edens2025spin,
  title={Spin and charge control of topological end states in chiral graphene nanoribbons on a 2D ferromagnet},
  author={Edens, Leonard and Romero-Lara, Francisco and Sai, Trisha and Biswas, Kalyan and Vilas-Varela, Manuel and Frederiksen, Thomas and Pe{\~n}a, Diego and Schulz, Fabian and Pascual, Jose Ignacio},
  journal={Advanced Materials},
  pages={e10753},
  year={2025},
  publisher={Wiley Online Library}
}

@article{kimouche2015ultra,
  title={Ultra-narrow metallic armchair graphene nanoribbons},
  author={Kimouche, Amina and Ervasti, Mikko M and Drost, Robert and Halonen, Simo and Harju, Ari and Joensuu, Pekka M and Sainio, Jani and Liljeroth, Peter},
  journal={Nature communications},
  volume={6},
  number={1},
  pages={10177},
  year={2015},
  publisher={Nature Publishing Group UK London}
}

@article{chen2013tuning,
  title={Tuning the band gap of graphene nanoribbons synthesized from molecular precursors},
  author={Chen, Yen-Chia and De Oteyza, Dimas G and Pedramrazi, Zahra and Chen, Chen and Fischer, Felix R and Crommie, Michael F},
  journal={ACS nano},
  volume={7},
  number={7},
  pages={6123--6128},
  year={2013},
  publisher={ACS Publications}
}

@article{kinikar2023surface,
  author    = {Kinikar, Amogh and Xu, Xiushang and Di Giovannantonio, Marco and Gr{\"o}ning, Oliver and Eimre, Kristjan and Pignedoli, Carlo A. and M{\"u}llen, Klaus and Narita, Akimitsu and Ruffieux, Pascal and Fasel, Roman},
  title     = {On-Surface Synthesis of Edge-Extended Zigzag Graphene Nanoribbons},
  journal   = {Advanced Materials},
  volume    = {35},
  number    = {48},
  pages     = {2306311},
  year      = {2023},
  publisher = {Wiley Online Library},
}

@article{zhang2026bottom,
  author    = {Zhang, Jian and Ghawri, Bhaskar and Dutta, Debopriya and Fasel, Roman and Calame, Michel and Borin Barin, Gabriela and Perrin, Mickael L.},
  title     = {Bottom-Up-Synthesized Graphene Nanoribbons for Nanoelectronics},
  journal   = {Nature Reviews Materials},
  pages     = {1--19},
  year      = {2026},
  publisher = {Nature Publishing Group},
}

@article{kinikar2024electronic,
  author    = {Kinikar, Amogh and Englmann, Thorsten G. and Di Giovannantonio, Marco and Bassi, Nicol{\`o} and Xiang, Feifei and Stolz, Samuel and Widmer, Roland and Borin Barin, Gabriela and Turco, Elia and Eimre, Kristjan and others},
  title     = {Electronic Decoupling and Hole-Doping of Graphene Nanoribbons on Metal Substrates by Chloride Intercalation},
  journal   = {ACS Nano},
  volume    = {18},
  number    = {26},
  pages     = {16622--16631},
  year      = {2024},
  publisher = {ACS Publications},
}

@article{blackwell2021spin,
  author    = {Blackwell, Raymond E. and Zhao, Fangzhou and Brooks, Erin and Zhu, Junmian and Piskun, Ilya and Wang, Shenkai and Delgado, Aidan and Lee, Yea-Lee and Louie, Steven G. and Fischer, Felix R.},
  title     = {Spin Splitting of Dopant Edge State in Magnetic Zigzag Graphene Nanoribbons},
  journal   = {Nature},
  volume    = {600},
  number    = {7890},
  pages     = {647--652},
  year      = {2021},
  publisher = {Nature Publishing Group UK London},
}

@article{wang2021cove,
  author    = {Wang, Xu and Ma, Ji and Zheng, Wenhao and Osella, Silvio and Arisnabarreta, Nicolas and Droste, Jorn and Serra, Gianluca and Ivasenko, Oleksandr and Lucotti, Andrea and Beljonne, David and others},
  title     = {Cove-Edged Graphene Nanoribbons with Incorporation of Periodic Zigzag-Edge Segments},
  journal   = {Journal of the American Chemical Society},
  volume    = {144},
  number    = {1},
  pages     = {228--235},
  year      = {2021},
  publisher = {ACS Publications},
}

@article{hwang2025optimized,
  author    = {Hwang, Jeong Ha and Bassi, Nicol{\`o} and Fadel, Mayada and Braun, Oliver and Dumslaff, Tim and Pignedoli, Carlo A. and Stiefel, Michael and Furrer, Roman and Hayashi, Hironobu and Yamada, Hiroko and others},
  title     = {Optimized Synthesis and Device Integration of Long 17-Atom-Wide Armchair Graphene Nanoribbons},
  journal   = {ACS Nano},
  year      = {2025},
  publisher = {ACS Publications},
}

@article{geagea2024growth,
  author    = {Geagea, Elie and Medina-Lopez, Daniel and Giovanelli, Luca and Nony, Laurent and Loppacher, Christian and Campidelli, St{\'e}phane and Clair, Sylvain},
  title     = {Growth Mechanism of Chevron Graphene Nanoribbons on (111)-Oriented Coinage Metal Surfaces},
  journal   = {The Journal of Physical Chemistry C},
  volume    = {128},
  number    = {21},
  pages     = {8601--8610},
  year      = {2024},
  publisher = {ACS Publications},
}

@article{cao2018tuning,
  author    = {Cao, Yun and Qi, Jing and Zhang, Yan-Fang and Huang, Li and Zheng, Qi and Lin, Xiao and Cheng, Zhihai and Zhang, Yu-Yang and Feng, Xinliang and Du, Shixuan and others},
  title     = {Tuning the Morphology of Chevron-Type Graphene Nanoribbons by Choice of Annealing Temperature},
  journal   = {Nano Research},
  volume    = {11},
  number    = {12},
  pages     = {6190--6196},
  year      = {2018},
  publisher = {Springer},
}

@article{lawrence2022circumventing,
  author    = {Lawrence, James and Berdonces-Layunta, Alejandro and Edalatmanesh, Shayan and Castro-Esteban, Jesus and Wang, Tao and Jimenez-Martin, Alejandro and de la Torre, Bruno and Castrillo-Bodero, Rodrigo and Angulo-Portugal, Paula and Mohammed, Mohammed S. G. and others},
  title     = {Circumventing the Stability Problems of Graphene Nanoribbon Zigzag Edges},
  journal   = {Nature Chemistry},
  volume    = {14},
  number    = {12},
  pages     = {1451--1458},
  year      = {2022},
  publisher = {Nature Publishing Group UK London},
}

@article{10.1039/c7nr07369k,
  author  = {Hempel, Marek and Lu, Ang-Yu and Hui, Fei and Kpulun, Tewa and Lanza, Mario and Harris, Gary and Palacios, Tom{\'a}s and Kong, Jing},
  title   = {Repeated Roll-to-Roll Transfer of Two-Dimensional Materials by Electrochemical Delamination},
  journal = {Nanoscale},
  year    = {2018},
  doi     = {10.1039/c7nr07369k},
}

@article{kaur2022fundamentals,
  author    = {Kaur, Gurkiran and Thimes, Rebekah L. and Camden, Jon P. and Jenkins, David M.},
  title     = {Fundamentals and Applications of N-Heterocyclic Carbene Functionalized Gold Surfaces and Nanoparticles},
  journal   = {Chemical Communications},
  volume    = {58},
  number    = {95},
  pages     = {13188--13197},
  year      = {2022},
  publisher = {Royal Society of Chemistry},
}

@article{trujillo2018using,
  author    = {Trujillo, Michael J. and Strausser, Shelby L. and Becca, Jeffrey C. and DeJesus, Joseph F. and Jensen, Lasse and Jenkins, David M. and Camden, Jon P.},
  title     = {Using SERS to Understand the Binding of N-Heterocyclic Carbenes to Gold Surfaces},
  journal   = {The Journal of Physical Chemistry Letters},
  volume    = {9},
  number    = {23},
  pages     = {6779--6785},
  year      = {2018},
  publisher = {ACS Publications},
}

@article{kim2017reduction,
  author    = {Kim, Hye Kyung and Hyla, Alexander S. and Winget, Paul and Li, Hong and Wyss, Chelsea M. and Jordan, Abraham J. and Larrain, Felipe A. and Sadighi, Joseph P. and Fuentes-Hernandez, Canek and Kippelen, Bernard and others},
  title     = {Reduction of the Work Function of Gold by N-Heterocyclic Carbenes},
  journal   = {Chemistry of Materials},
  volume    = {29},
  number    = {8},
  pages     = {3403--3411},
  year      = {2017},
  publisher = {ACS Publications},
}

@article{daukiya2018highly,
  author    = {Daukiya, Lakshya and Nair, M. N. and Hajjar-Garreau, Samar and Vonau, Francois and Aubel, Dominique and Bubendorff, Jean-Luc and Cranney, Marion and Denys, Emmanuel and Florentin, Alban and Reiter, G{\"u}nter and others},
  title     = {Highly n-Doped Graphene Generated Through Intercalated Terbium Atoms},
  journal   = {Physical Review B},
  volume    = {97},
  number    = {3},
  pages     = {035309},
  year      = {2018},
  publisher = {APS},
}

@article{10.1021/acsnano.0c01733,
  author  = {Krzykawska, Anna and Wr{\'o}bel, Mateusz and Kozie{\l}, Krzysztof and Cyganik, Piotr},
  title   = {N-Heterocyclic Carbenes for the Self-Assembly of Thin and Highly Insulating Monolayers With High Quality and Stability},
  journal = {ACS Nano},
  year    = {2020},
  doi     = {10.1021/acsnano.0c01733},
}

@article{10.1002/cphc.201900313,
  author  = {Talirz, Leopold and S{\"o}de, Hajo and Kawai, Shigeki and Ruffieux, Pascal and Meyer, Ernst and Feng, Xinliang and M{\"u}llen, Klaus and Fasel, Roman and Pignedoli, Carlo A. and Passerone, Daniele},
  title   = {Band Gap of Atomically Precise Graphene Nanoribbons as a Function of Ribbon Length and Termination},
  journal = {ChemPhysChem},
  year    = {2019},
  doi     = {10.1002/cphc.201900313},
}

@article{10.1002/adfm.202103798,
  author  = {Mutlu, Zafer and Jacobse, Peter H. and McCurdy, Ryan D. and Llinas, Juan Pablo and Lin, Yuxuan and Veber, Gregory and Fischer, Felix R. and Crommie, Michael F. and Bokor, Jeffrey},
  title   = {Bottom-Up Synthesized Nanoporous Graphene Transistors},
  journal = {Advanced Functional Materials},
  year    = {2021},
  doi     = {10.1002/adfm.202103798},
}

@article{yang2007quasiparticle,
  author    = {Yang, Li and Park, Cheol-Hwan and Son, Young-Woo and Cohen, Marvin L. and Louie, Steven G.},
  title     = {Quasiparticle Energies and Band Gaps in Graphene Nanoribbons},
  journal   = {Physical Review Letters},
  volume    = {99},
  number    = {18},
  pages     = {186801},
  year      = {2007},
  publisher = {APS},
}

@article{merino2017width,
  author    = {Merino-D{\'\i}ez, N{\'e}stor and Garcia-Lekue, Aran and Carbonell-Sanrom{\`a}, Eduard and Li, Jingcheng and Corso, Martina and Colazzo, Luciano and Sedona, Francesco and S{\'a}nchez-Portal, Daniel and Pascual, Jose I. and De Oteyza, Dimas G.},
  title     = {Width-Dependent Band Gap in Armchair Graphene Nanoribbons Reveals Fermi Level Pinning on {Au(111)}},
  journal   = {ACS Nano},
  volume    = {11},
  number    = {11},
  pages     = {11661--11668},
  year      = {2017},
  publisher = {ACS Publications},
}

@article{cai2010atomically,
  author    = {Cai, Jinming and Ruffieux, Pascal and Jaafar, Rached and Bieri, Marco and Braun, Thomas and Blankenburg, Stephan and Muoth, Matthias and Seitsonen, Ari P. and Saleh, Moussa and Feng, Xinliang and others},
  title     = {Atomically Precise Bottom-Up Fabrication of Graphene Nanoribbons},
  journal   = {Nature},
  volume    = {466},
  number    = {7305},
  pages     = {470--473},
  year      = {2010},
  publisher = {Nature Publishing Group UK London},
}

@article{das2008monitoring,
  author    = {Das, Anindya and Pisana, Simone and Chakraborty, Biswanath and Piscanec, Stefano and Saha, Srijan K. and Waghmare, Umesh V. and Novoselov, Konstantin S. and Krishnamurthy, Hulikal R. and Geim, Andre K. and Ferrari, Andrea C. and others},
  title     = {Monitoring Dopants by Raman Scattering in an Electrochemically Top-Gated Graphene Transistor},
  journal   = {Nature Nanotechnology},
  volume    = {3},
  number    = {4},
  pages     = {210--215},
  year      = {2008},
  publisher = {Nature Publishing Group UK London},
}

@article{das2008raman,
  author    = {Das, Anindya and Chakraborty, Biswanath and Sood, Ajay Kumar},
  title     = {Raman Spectroscopy of Graphene on Different Substrates and Influence of Defects},
  journal   = {Bulletin of Materials Science},
  volume    = {31},
  pages     = {579--584},
  year      = {2008},
  publisher = {Springer},
}

@article{eckmann2012probing,
  author    = {Eckmann, Axel and Felten, Alexandre and Mishchenko, Artem and Britnell, Liam and Krupke, Ralph and Novoselov, Kostya S. and Casiraghi, Cinzia},
  title     = {Probing the Nature of Defects in Graphene by Raman Spectroscopy},
  journal   = {Nano Letters},
  volume    = {12},
  number    = {8},
  pages     = {3925--3930},
  year      = {2012},
  publisher = {ACS Publications},
}

@article{wang2016giant,
  author    = {Wang, Shiyong and Talirz, Leopold and Pignedoli, Carlo A. and Feng, Xinliang and M{\"u}llen, Klaus and Fasel, Roman and Ruffieux, Pascal},
  title     = {Giant Edge State Splitting at Atomically Precise Graphene Zigzag Edges},
  journal   = {Nature Communications},
  volume    = {7},
  number    = {1},
  pages     = {11507},
  year      = {2016},
  publisher = {Nature Publishing Group UK London},
}

@article{doi:10.1021/acsanm.5c02753,
  author  = {Kinikar, Amogh and Xiang, Feifei and Palomino-Ruiz, Lucia and Lu, Li-Syuan and Dong, Chengye and Gu, Yanwei and Darawish, Rimah and Ammerman, Eve and Gr{\"o}ning, Oliver and M{\"u}llen, Klaus and Fasel, Roman and Robinson, Joshua A. and Ruffieux, Pascal and Schuler, Bruno and Borin Barin, Gabriela},
  title   = {Atomic-Scale Imaging of Transferred Graphene Nanoribbons for Nanoelectronic Device Integration},
  journal = {ACS Applied Nano Materials},
  volume  = {8},
  number  = {33},
  pages   = {16457--16464},
  year    = {2025},
  doi     = {10.1021/acsanm.5c02753},
  url     = {https://doi.org/10.1021/acsanm.5c02753},
  eprint  = {https://doi.org/10.1021/acsanm.5c02753},
}

@article{senkovskiy2017making,
  author    = {Senkovskiy, B. V. and Pfeiffer, M. and Alavi, S. K. and Bliesener, A. and Zhu, J. and Michel, S. and Fedorov, A. V. and German, R. and Hertel, D. and Haberer, D. and others},
  title     = {Making Graphene Nanoribbons Photoluminescent},
  journal   = {Nano Letters},
  volume    = {17},
  number    = {7},
  pages     = {4029--4037},
  year      = {2017},
  publisher = {ACS Publications},
}

@article{deniz2017revealing,
  author    = {Deniz, Okan and S{\'a}nchez-S{\'a}nchez, Carlos and Dumslaff, Tim and Feng, Xinliang and Narita, Akimitsu and M{\"u}llen, Klaus and Kharche, Neerav and Meunier, Vincent and Fasel, Roman and Ruffieux, Pascal},
  title     = {Revealing the Electronic Structure of Silicon Intercalated Armchair Graphene Nanoribbons by Scanning Tunneling Spectroscopy},
  journal   = {Nano Letters},
  volume    = {17},
  number    = {4},
  pages     = {2197--2203},
  year      = {2017},
  publisher = {ACS Publications},
}

@article{bakker2018elucidating,
  author    = {Bakker, Anne and Timmer, Alexander and Kolodzeiski, Elena and Freitag, Matthias and Gao, Hong Ying and M{\"o}nig, Harry and Amirjalayer, Saeed and Glorius, Frank and Fuchs, Harald},
  title     = {Elucidating the Binding Modes of N-Heterocyclic Carbenes on a Gold Surface},
  journal   = {Journal of the American Chemical Society},
  volume    = {140},
  number    = {38},
  pages     = {11889--11892},
  year      = {2018},
  publisher = {ACS Publications},
}

@article{groning2018engineering,
  author    = {Gr{\"o}ning, Oliver and Wang, Shiyong and Yao, Xuelin and Pignedoli, Carlo A. and Borin Barin, Gabriela and Daniels, Colin and Cupo, Andrew and Meunier, Vincent and Feng, Xinliang and Narita, Akimitsu and M{\"u}llen, Klaus and Ruffieux, Pascal and Fasel, Roman},
  title     = {Engineering of Robust Topological Quantum Phases in Graphene Nanoribbons},
  journal   = {Nature},
  volume    = {560},
  number    = {7717},
  pages     = {209--213},
  year      = {2018},
  publisher = {Nature Publishing Group UK London},
}

@article{talirz2017surface,
  author    = {Talirz, Leopold and Sode, Hajo and Dumslaff, Tim and Wang, Shiyong and Sanchez-Valencia, Juan Ramon and Liu, Jia and Shinde, Prashant and Pignedoli, Carlo A. and Liang, Liangbo and Meunier, Vincent and others},
  title     = {On-Surface Synthesis and Characterization of 9-Atom Wide Armchair Graphene Nanoribbons},
  journal   = {ACS Nano},
  volume    = {11},
  number    = {2},
  pages     = {1380--1388},
  year      = {2017},
  publisher = {ACS Publications},
}

@article{zhukhovitskiy2015carbene,
  author    = {Zhukhovitskiy, Aleksandr V. and MacLeod, Michelle J. and Johnson, Jeremiah A.},
  title     = {Carbene Ligands in Surface Chemistry: From Stabilization of Discrete Elemental Allotropes to Modification of Nanoscale and Bulk Substrates},
  journal   = {Chemical Reviews},
  volume    = {115},
  number    = {20},
  pages     = {11503--11532},
  year      = {2015},
  publisher = {ACS Publications},
}

@article{larrea2017,
  author    = {Larrea, Christian R. and Baddeley, Christopher J. and Narouz, Mina R. and Mosey, Nicholas J. and Horton, J. Hugh and Crudden, Cathleen M.},
  title     = {N-Heterocyclic Carbene Self-Assembled Monolayers on Copper and Gold: Dramatic Effect of Wingtip Groups on Binding, Orientation and Assembly},
  journal   = {ChemPhysChem},
  volume    = {18},
  number    = {24},
  pages     = {3536--3539},
  year      = {2017},
  publisher = {Wiley Online Library},
}

@article{inayeh2021self,
  author    = {Inayeh, Alex and Groome, Ryan R. K. and Singh, Ishwar and Veinot, Alex J. and de Lima, Felipe Crasto and Miwa, Roberto H. and Crudden, Cathleen M. and McLean, Alastair B.},
  title     = {Self-Assembly of N-Heterocyclic Carbenes on {Au(111)}},
  journal   = {Nature Communications},
  volume    = {12},
  number    = {1},
  pages     = {4034},
  year      = {2021},
  publisher = {Nature Publishing Group UK London},
}

@article{borin2019surface,
  author    = {Borin Barin, Gabriela and Fairbrother, Andrew and Rotach, Lukas and Bayle, Maxime and Paillet, Matthieu and Liang, Liangbo and Meunier, Vincent and Hauert, Roland and Dumslaff, Tim and Narita, Akimitsu and others},
  title     = {Surface-Synthesized Graphene Nanoribbons for Room Temperature Switching Devices: Substrate Transfer and ex Situ Characterization},
  journal   = {ACS Applied Nano Materials},
  volume    = {2},
  number    = {4},
  pages     = {2184--2192},
  year      = {2019},
  publisher = {ACS Publications},
}

@article{chen2014mechanistic,
  title={Mechanistic Study of a Switch in the Regioselectivity of Hydroheteroarylation of Styrene Catalyzed by Bimetallic Ni--Al through C--H Activation},
  author={Chen, Wen-Ching and Lai, Ying-Chieh and Shih, Wei-Chun and Yu, Ming-Shiuan and Yap, Glenn PA and Ong, Tiow-Gan},
  journal={Chemistry--A European Journal},
  volume={20},
  number={26},
  pages={8099--8105},
  year={2014},
  publisher={Wiley Online Library}
}

@article{overbeck2019universal,
  author    = {Overbeck, Jan and Barin, Gabriela Borin and Daniels, Colin and Perrin, Mickael L. and Braun, Oliver and Sun, Qiang and Darawish, Rimah and De Luca, Marta and Wang, Xiao-Ye and Dumslaff, Tim and others},
  title     = {A Universal Length-Dependent Vibrational Mode in Graphene Nanoribbons},
  journal   = {ACS Nano},
  volume    = {13},
  number    = {11},
  pages     = {13083--13091},
  year      = {2019},
  publisher = {ACS Publications},
}

@article{son2006energy,
  author    = {Son, Young-Woo and Cohen, Marvin L. and Louie, Steven G.},
  title     = {Energy Gaps in Graphene Nanoribbons},
  journal   = {Physical Review Letters},
  volume    = {97},
  number    = {21},
  pages     = {216803},
  year      = {2006},
  publisher = {APS},
}

@article{yazyev2013guide,
  author    = {Yazyev, Oleg V.},
  title     = {A Guide to the Design of Electronic Properties of Graphene Nanoribbons},
  journal   = {Accounts of Chemical Research},
  volume    = {46},
  number    = {10},
  pages     = {2319--2328},
  year      = {2013},
  publisher = {ACS Publications},
}

@article{yano2019quest,
  author    = {Yano, Yuuta and Mitoma, Nobuhiko and Ito, Hideto and Itami, Kenichiro},
  title     = {A Quest for Structurally Uniform Graphene Nanoribbons: Synthesis, Properties, and Applications},
  journal   = {The Journal of Organic Chemistry},
  volume    = {85},
  number    = {1},
  pages     = {4--33},
  year      = {2019},
  publisher = {ACS Publications},
}

@article{ruffieux2016surface,
  author    = {Ruffieux, Pascal and Wang, Shiyong and Yang, Bo and S{\'a}nchez-S{\'a}nchez, Carlos and Liu, Jia and Dienel, Thomas and Talirz, Leopold and Shinde, Prashant and Pignedoli, Carlo A. and Passerone, Daniele and others},
  title     = {On-Surface Synthesis of Graphene Nanoribbons with Zigzag Edge Topology},
  journal   = {Nature},
  volume    = {531},
  number    = {7595},
  pages     = {489--492},
  year      = {2016},
  publisher = {Nature Publishing Group UK London},
}

@article{richter2020charge,
  author    = {Richter, Nils and Chen, Zongping and Tries, Alexander and Prechtl, Thorsten and Narita, Akimitsu and M{\"u}llen, Klaus and Asadi, Kamal and Bonn, Mischa and Kl{\"a}ui, Mathias},
  title     = {Charge Transport Mechanism in Networks of Armchair Graphene Nanoribbons},
  journal   = {Scientific Reports},
  volume    = {10},
  number    = {1},
  pages     = {1988},
  year      = {2020},
  publisher = {Nature Publishing Group UK London},
}

@article{ruffieux2012electronic,
  author    = {Ruffieux, Pascal and Cai, Jinming and Plumb, Nicholas C. and Patthey, Luc and Prezzi, Deborah and Ferretti, Andrea and Molinari, Elisa and Feng, Xinliang and M{\"u}llen, Klaus and Pignedoli, Carlo A. and others},
  title     = {Electronic Structure of Atomically Precise Graphene Nanoribbons},
  journal   = {ACS Nano},
  volume    = {6},
  number    = {8},
  pages     = {6930--6935},
  year      = {2012},
  publisher = {ACS Publications},
}

@article{ohtomo2016etchant,
  author    = {Ohtomo, Manabu and Sekine, Yoshiaki and Wang, Shengnan and Hibino, Hiroki and Yamamoto, Hideki},
  title     = {Etchant-Free Graphene Transfer Using Facile Intercalation of Alkanethiol Self-Assembled Molecules at Graphene/Metal Interfaces},
  journal   = {Nanoscale},
  volume    = {8},
  number    = {22},
  pages     = {11503--11510},
  year      = {2016},
  publisher = {Royal Society of Chemistry},
}

@article{arnay2025tuning,
  author    = {Arnay, Ic{\'\i}ar and Guedeja-Marr{\'o}n, Alejandra and Gud{\'\i}n, Adri{\'a}n and Guerrero, Rub{\'e}n and Diez, Jos{\'e} Manuel and Anad{\'o}n, Alberto and Varela, Mar{\'\i}a and Foerster, Michael and Ni{\~n}o, Miguel {\'A}ngel and Camarero, Julio and others},
  title     = {Tuning Work Function in Graphene by Thermally Assisted Ferromagnetic Metal Intercalation},
  journal   = {Applied Surface Science},
  pages     = {163733},
  year      = {2025},
  publisher = {Elsevier},
}

@article{que2020surface,
  author    = {Que, Yande and Liu, Bin and Zhuang, Yuan and Xu, Chaoqiang and Wang, Kedong and Xiao, Xudong},
  title     = {On-Surface Synthesis of Graphene Nanoribbons on Two-Dimensional Rare Earth--Gold Intermetallic Compounds},
  journal   = {The Journal of Physical Chemistry Letters},
  volume    = {11},
  number    = {13},
  pages     = {5044--5050},
  year      = {2020},
  publisher = {ACS Publications},
}

@article{verzhbitskiy2016raman,
  author    = {Verzhbitskiy, Ivan A. and De Corato, Marzio and Ruini, Alice and Molinari, Elisa and Narita, Akimitsu and Hu, Yunbin and Schwab, Matthias G. and Bruna, Matteo and Yoon, Duhee and Milana, Silvia and others},
  title     = {Raman Fingerprints of Atomically Precise Graphene Nanoribbons},
  journal   = {Nano Letters},
  volume    = {16},
  number    = {6},
  pages     = {3442--3447},
  year      = {2016},
  publisher = {ACS Publications},
}

@article{hell2018resonance,
  author    = {Hell, Martin G. and Ehlen, Niels and Senkovskiy, Boris V. and Hasdeo, Eddwi H. and Fedorov, Alexander and Dombrowski, Daniela and Busse, Carsten and Michely, Thomas and Di Santo, Giovanni and Petaccia, Luca and others},
  title     = {Resonance Raman Spectrum of Doped Epitaxial Graphene at the Lifshitz Transition},
  journal   = {Nano Letters},
  volume    = {18},
  number    = {9},
  pages     = {6045--6056},
  year      = {2018},
  publisher = {ACS Publications},
}

@article{barin2023surface,
  title={On-surface synthesis and characterization of teranthene and hexanthene: ultrashort graphene nanoribbons with mixed armchair and zigzag edges},
  author={Barin, Gabriela Borin and Di Giovannantonio, Marco and Lohr, Thorsten G and Mishra, Shantanu and Kinikar, Amogh and Perrin, Mickael L and Overbeck, Jan and Calame, Michel and Feng, Xinliang and Fasel, Roman and others},
  journal={Nanoscale},
  volume={15},
  number={41},
  pages={16766--16774},
  year={2023},
  publisher={Royal Society of Chemistry}
}

@article{bonini2007phonon,
  author    = {Bonini, Nicola and Lazzeri, Michele and Marzari, Nicola and Mauri, Francesco},
  title     = {Phonon Anharmonicities in Graphite and Graphene},
  journal   = {Physical Review Letters},
  volume    = {99},
  number    = {17},
  pages     = {176802},
  year      = {2007},
  publisher = {APS},
}

@article{basko2009electron,
  author    = {Basko, D. M. and Piscanec, S. and Ferrari, A. C.},
  title     = {Electron-Electron Interactions and Doping Dependence of the Two-Phonon Raman Intensity in Graphene},
  journal   = {Physical Review B},
  volume    = {80},
  number    = {16},
  pages     = {165413},
  year      = {2009},
  publisher = {APS},
}

@article{crudden2014ultrastable,
  author    = {Crudden, Cathleen M. and Horton, J. Hugh and Ebralidze, Iraklii I. and Zenkina, Olena V. and McLean, Alastair B. and Drevniok, Benedict and She, Zhe and Kraatz, Heinz-Bernhard and Mosey, Nicholas J. and Seki, Tomohiro and others},
  title     = {Ultra Stable Self-Assembled Monolayers of N-Heterocyclic Carbenes on Gold},
  journal   = {Nature Chemistry},
  volume    = {6},
  number    = {5},
  pages     = {409--414},
  year      = {2014},
  publisher = {Nature Publishing Group UK London},
}

@article{engel2017new,
  author    = {Engel, Sabrina and Fritz, Eva-Corinna and Ravoo, Bart Jan},
  title     = {New Trends in the Functionalization of Metallic Gold: From Organosulfur Ligands to N-Heterocyclic Carbenes},
  journal   = {Chemical Society Reviews},
  volume    = {46},
  number    = {8},
  pages     = {2057--2075},
  year      = {2017},
  publisher = {Royal Society of Chemistry},
}

@inproceedings{casiraghi2017raman,
  title={Raman spectroscopy of graphene nanoribbons: A review},
  author={Casiraghi, C and Prezzi, D},
  booktitle={GraphITA: Selected papers from the Workshop on Synthesis, Characterization and Technological Exploitation of Graphene and 2D Materials Beyond Graphene},
  pages={19--30},
  year={2017},
  organization={Springer}
}

@article{guo2022phonon,
  title={Phonon anharmonicities in 7-armchair graphene nanoribbons},
  author={Guo, Xiao and Tian, Qiwei and Wang, Yongsong and Liu, Jinxin and Jia, Guiping and Dou, Weidong and Song, Fei and Zhang, Lijie and Qin, Zhihui and Huang, Han},
  journal={Carbon},
  volume={190},
  pages={312--318},
  year={2022},
  publisher={Elsevier}
}

@article{brede2023detecting,
  author    = {Brede, Jens and Merino-D{\'\i}ez, Nestor and Berdonces-Layunta, Alejandro and Sanz, Sof{\'\i}a and Dom{\'\i}nguez-Celorrio, Amelia and Lobo-Checa, Jorge and Vilas-Varela, Manuel and Pe{\~n}a, Diego and Frederiksen, Thomas and Pascual, Jos{\'e} I. and others},
  title     = {Detecting the Spin-Polarization of Edge States in Graphene Nanoribbons},
  journal   = {Nature Communications},
  volume    = {14},
  number    = {1},
  pages     = {6677},
  year      = {2023},
  publisher = {Nature Publishing Group UK London},
}

@article{overbeck2019optimized,
  author    = {Overbeck, Jan and Borin Barin, Gabriela and Daniels, Colin and Perrin, Mickael L. and Liang, Liangbo and Braun, Oliver and Darawish, Rimah and Burkhardt, Bryanna and Dumslaff, Tim and Wang, Xiao-Ye and others},
  title     = {Optimized Substrates and Measurement Approaches for Raman Spectroscopy of Graphene Nanoribbons},
  journal   = {Physica Status Solidi (B)},
  volume    = {256},
  number    = {12},
  pages     = {1900343},
  year      = {2019},
  publisher = {Wiley Online Library},
}

@article{rizzo2018topological,
  author    = {Rizzo, Daniel J. and Veber, Gregory and Cao, Ting and Bronner, Christopher and Chen, Ting and Zhao, Fangzhou and Rodriguez, Henry and Louie, Steven G. and Crommie, Michael F. and Fischer, Felix R.},
  title     = {Topological Band Engineering of Graphene Nanoribbons},
  journal   = {Nature},
  volume    = {560},
  number    = {7717},
  pages     = {204--208},
  year      = {2018},
  publisher = {Nature Publishing Group UK London},
}

@article{bae2010roll,
  author    = {Bae, Sukang and Kim, Hyeongkeun and Lee, Youngbin and Xu, Xiangfan and Park, Jae-Sung and Zheng, Yi and Balakrishnan, Jayakumar and Lei, Tian and Kim, Hye and Song, Young Il and others},
  title     = {Roll-to-Roll Production of 30-Inch Graphene Films for Transparent Electrodes},
  journal   = {Nature Nanotechnology},
  volume    = {5},
  number    = {8},
  pages     = {574--578},
  year      = {2010},
  publisher = {Nature Publishing Group UK London},
}

@article{houtsma2021atomically,
  author    = {Houtsma, R. S. Koen and de la Rie, Joris and St{\"o}hr, Meike},
  title     = {Atomically Precise Graphene Nanoribbons: Interplay of Structural and Electronic Properties},
  journal   = {Chemical Society Reviews},
  volume    = {50},
  number    = {11},
  pages     = {6541--6568},
  year      = {2021},
  publisher = {Royal Society of Chemistry},
}

@article{aiidalab,
  author    = {Yakutovich, Aliaksandr V. and Eimre, Kristjan and Sch{\"u}tt, Ole and Talirz, Leopold and Adorf, Carl S. and Andersen, Casper W. and Ditler, Edward and Du, Dou and Passerone, Daniele and Smit, Berend and others},
  title     = {{AiiDAlab}---an ecosystem for developing, executing, and sharing scientific workflows},
  journal   = {Computational Materials Science},
  volume    = {188},
  pages     = {110165},
  year      = {2021},
  publisher = {Elsevier},
}

@article{aiida,
  author    = {Pizzi, Giovanni and Cepellotti, Andrea and Sabatini, Riccardo and Marzari, Nicola and Kozinsky, Boris},
  title     = {{AiiDA}: automated interactive infrastructure and database for computational science},
  journal   = {Computational Materials Science},
  volume    = {111},
  pages     = {218--230},
  year      = {2016},
  publisher = {Elsevier},
}

@article{pbe,
  author    = {Perdew, John P. and Burke, Kieron and Ernzerhof, Matthias},
  title     = {Generalized Gradient Approximation Made Simple},
  journal   = {Physical Review Letters},
  volume    = {77},
  number    = {18},
  pages     = {3865},
  year      = {1996},
  publisher = {APS},
}

@article{vdwd3,
  author    = {Goedecker, Stefan and Teter, Michael and Hutter, J{\"u}rg},
  title     = {Separable Dual-Space Gaussian Pseudopotentials},
  journal   = {Physical Review B},
  volume    = {54},
  number    = {3},
  pages     = {1703},
  year      = {1996},
  publisher = {APS},
}

@article{basis,
  author    = {VandeVondele, Joost and Hutter, J{\"u}rg},
  title     = {Gaussian Basis Sets for Accurate Calculations on Molecular Systems in Gas and Condensed Phases},
  journal   = {The Journal of Chemical Physics},
  volume    = {127},
  number    = {11},
  pages     = {114105},
  year      = {2007},
  publisher = {American Institute of Physics},
}

@article{cp2k,
  author    = {Hutter, J{\"u}rg and Iannuzzi, Marcella and Schiffmann, Florian and VandeVondele, Joost},
  title     = {{CP2K}: atomistic simulations of condensed matter systems},
  journal   = {Wiley Interdisciplinary Reviews: Computational Molecular Science},
  volume    = {4},
  number    = {1},
  pages     = {15--25},
  year      = {2014},
  publisher = {Wiley Online Library},
}

@Inbook{Tersoff1993,
  author    = {Tersoff, J. and Hamann, D. R.},
  editor    = {Neddermeyer, H.},
  title     = {Theory of the scanning tunneling microscope},
  bookTitle = {Scanning Tunneling Microscopy},
  year      = {1993},
  publisher = {Springer Netherlands},
  address   = {Dordrecht},
  pages     = {59--67},
  isbn      = {978-94-011-1812-5},
  doi       = {10.1007/978-94-011-1812-5_5},
  url       = {https://doi.org/10.1007/978-94-011-1812-5_5},
}

@article{willand2013norm,
  author    = {Willand, Alex and Kvashnin, Yaroslav O. and Genovese, Luigi and V{\'a}zquez-Mayagoitia, {\'A}lvaro and Deb, Arpan Krishna and Sadeghi, Ali and Deutsch, Thierry and Goedecker, Stefan},
  title     = {Norm-Conserving Pseudopotentials with Chemical Accuracy Compared to All-Electron Calculations},
  journal   = {The Journal of Chemical Physics},
  volume    = {138},
  number    = {10},
  year      = {2013},
  publisher = {AIP Publishing},
}

@article{krzykawska2020n,
  title={N-heterocyclic carbenes for the self-assembly of thin and highly insulating monolayers with high quality and stability},
  author={Krzykawska, Anna and Wrobel, Mateusz and Kozie{\l}, Krzysztof and Cyganik, Piotr},
  journal={ACS nano},
  volume={14},
  number={5},
  pages={6043--6057},
  year={2020},
  publisher={ACS Publications}
}

@article{grimme2010consistent,
  title={A consistent and accurate ab initio parametrization of density functional dispersion correction (DFT-D) for the 94 elements H-Pu},
  author={Grimme, Stefan and Antony, Jens and Ehrlich, Stephan and Krieg, Helge},
  journal={The Journal of chemical physics},
  volume={132},
  number={15},
  year={2010},
  publisher={AIP Publishing}
}

@article{crudden2016simple,
  title={Simple direct formation of self-assembled N-heterocyclic carbene monolayers on gold and their application in biosensing},
  author={Crudden, Cathleen M and Horton, J Hugh and Narouz, Mina R and Li, Zhijun and Smith, Christene A and Munro, Kim and Baddeley, Christopher J and Larrea, Christian R and Drevniok, Benedict and Thanabalasingam, Bheeshmon and others},
  journal={Nature Communications},
  volume={7},
  number={1},
  pages={12654},
  year={2016},
  publisher={Nature Publishing Group UK London}
}

@article{narita2019solution,
  title={Solution and on-surface synthesis of structurally defined graphene nanoribbons as a new family of semiconductors},
  author={Narita, Akimitsu and Chen, Zongping and Chen, Qiang and M{\"u}llen, Klaus},
  journal={Chemical science},
  volume={10},
  number={4},
  pages={964--975},
  year={2019},
  publisher={Royal Society of Chemistry}
}

\clearpage
\section*{Supporting Information}
\addcontentsline{toc}{section}{Supporting Information}

\setcounter{secnumdepth}{2}
\setcounter{section}{0}
\renewcommand{\thesection}{S\arabic{section}}
\renewcommand{\thesubsection}{S\arabic{section}.\arabic{subsection}}

\setcounter{figure}{0}
\renewcommand{\thefigure}{S\arabic{figure}}
\renewcommand{\theHfigure}{S\arabic{figure}} 

\setcounter{table}{0}
\renewcommand{\thetable}{S\arabic{table}}
\renewcommand{\theHtable}{S\arabic{table}}   

\setcounter{equation}{0}
\renewcommand{\theequation}{S\arabic{equation}}
\renewcommand{\theHequation}{S\arabic{equation}} 

\renewcommand{\thesection}{S\arabic{section}}

\section{Detailed synthetic procedure and characterization data of N-heterocyclic carbenes (NHCs)}
\label{sec:synthesis}

\begin{figure}[h!]
    \centering
    \includegraphics[width=0.9\linewidth]{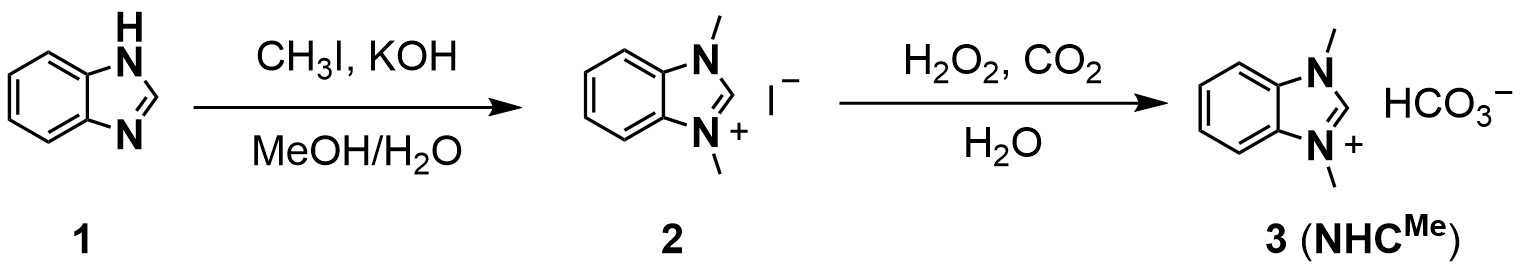}
\end{figure}

\subsection*{Synthesis of 1,3-dimethyl-1H-benzo[d]imidazol-3-ium iodide (\textbf{2})}

A mixture of benzimidazole (236.3 mg, 2.0 mmol) in 2 N KOH solution (5.8 mL in H$_2$O) and MeOH (1.2 mL) was added to a 25 mL round-bottom flask fitted with a condenser. Methyl iodide (1.40 mL, 22.4 mmol) was added dropwise and the mixture was stirred at 45~$^\circ$C for 12 h. The reaction yielded 1,3-dimethylbenzoimidazolium iodide (372.6 mg, 68\%) as a white powder, which was used in the next step without further purification.

\subsection*{Synthesis of 1,3-dimethyl-1H-benzo[d]imidazol-3-ium hydrogencarbonate (\textbf{3}, NHC\textsuperscript{Me})}

A 25 mL round-bottom flask, capped with a rubber septum and equipped with a vent needle and a glass pipette for the addition of gaseous CO$_2$, was charged with compound \textbf{2} (274.0 mg, 1.0 mmol) in deionized water (10 mL). Carbon dioxide was bubbled through the stirred solution for 2--3 minutes. Subsequently, hydrogen peroxide (150 $\mu$L, 30\%, diluted in 1 mL water) was injected via syringe, resulting in an immediate dark violet coloration indicative of free iodine formation.

Vigorous bubbling of CO$_2$ was maintained for 35 minutes under continuous stirring. The precipitated iodine was removed by filtration and washed with deionized water. The filtrate was exposed to a stream of air for 24 h to remove residual water. The resulting solid was dried under vacuum at ambient temperature to afford a yellowish powder, which was triturated with acetone three times to yield 103.5 mg (49.7\%) of compound \textbf{3} (NHC\textsuperscript{Me}) as a white powder.

$^1$H NMR (300 MHz, D$_2$O): 7.89--7.85 (m, 2H), 7.76--7.71 (m, 2H), 4.11 (s, 6H), consistent with literature.\cite{krzykawska2020n}

Note: The proton at position 2 and the HCO$_3^-$ protons were not observed due to rapid exchange with the deuterated solvent on the NMR time scale.

\begin{figure}[h!]
    \centering
    \includegraphics[width=0.9\linewidth]{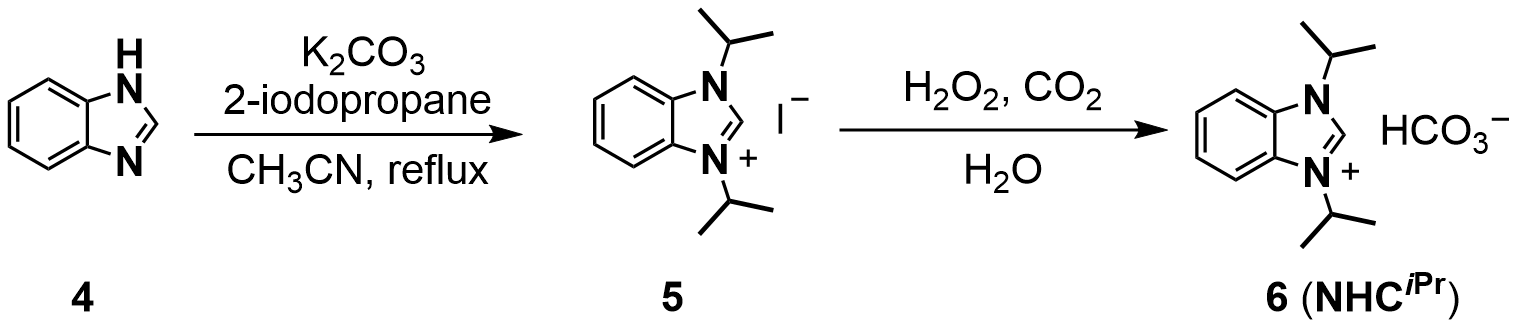}
\end{figure}

\subsection*{Synthesis of 1,3-diisopropyl-1H-benzo[d]imidazol-3-ium iodide (\textbf{5})}

In a 25 mL sealed round-bottom flask, benzimidazole (2.00 g, 16.93 mmol) and K$_2$CO$_3$ (2.57 g, 18.62 mmol) were dissolved in CH$_3$CN (10 mL). After stirring for 1 h, 2-iodopropane (8.63 g, 50.79 mmol) was added. The reaction mixture was refluxed for 2 days.

The mixture was quenched with water and extracted with CH$_2$Cl$_2$ (three times). The combined organic layers were dried over MgSO$_4$, filtered, and concentrated under reduced pressure. The crude product was washed with ethyl acetate and dried under vacuum to afford compound \textbf{5} (5.33 g, 95\%) as a solid, which was used in the next step without further purification.

\subsection*{Synthesis of 1,3-diisopropyl-1H-benzo[d]imidazol-3-ium hydrogencarbonate (\textbf{6}, NHC\textsuperscript{\textit{i}Pr})}

Following a procedure analogous to that used for compound \textbf{3}, compound \textbf{6} (NHC\textsuperscript{\textit{i}Pr}) was obtained as a white solid in 45\% yield from compound \textbf{5}.

$^1$H NMR (300 MHz, D$_2$O): 7.97--7.91 (m, 2H), 7.71--7.65 (m, 2H), 5.08--4.95 (m, 2H), 1.69 (s, 6H), 1.67 (s, 6H), consistent with literature.\cite{crudden2016simple}

Note: The proton at position 2 and the HCO$_3^-$ protons were not observed due to rapid exchange with the deuterated solvent on the NMR time scale.

Compounds 2 and 5 were prepared according to literature procedures.\cite{chen2014mechanistic}  
Compounds 3 and 6 were prepared according to literature procedures.\cite{krzykawska2020n,crudden2016simple}

\begin{figure}
    \centering
    \includegraphics[width=0.8\linewidth]{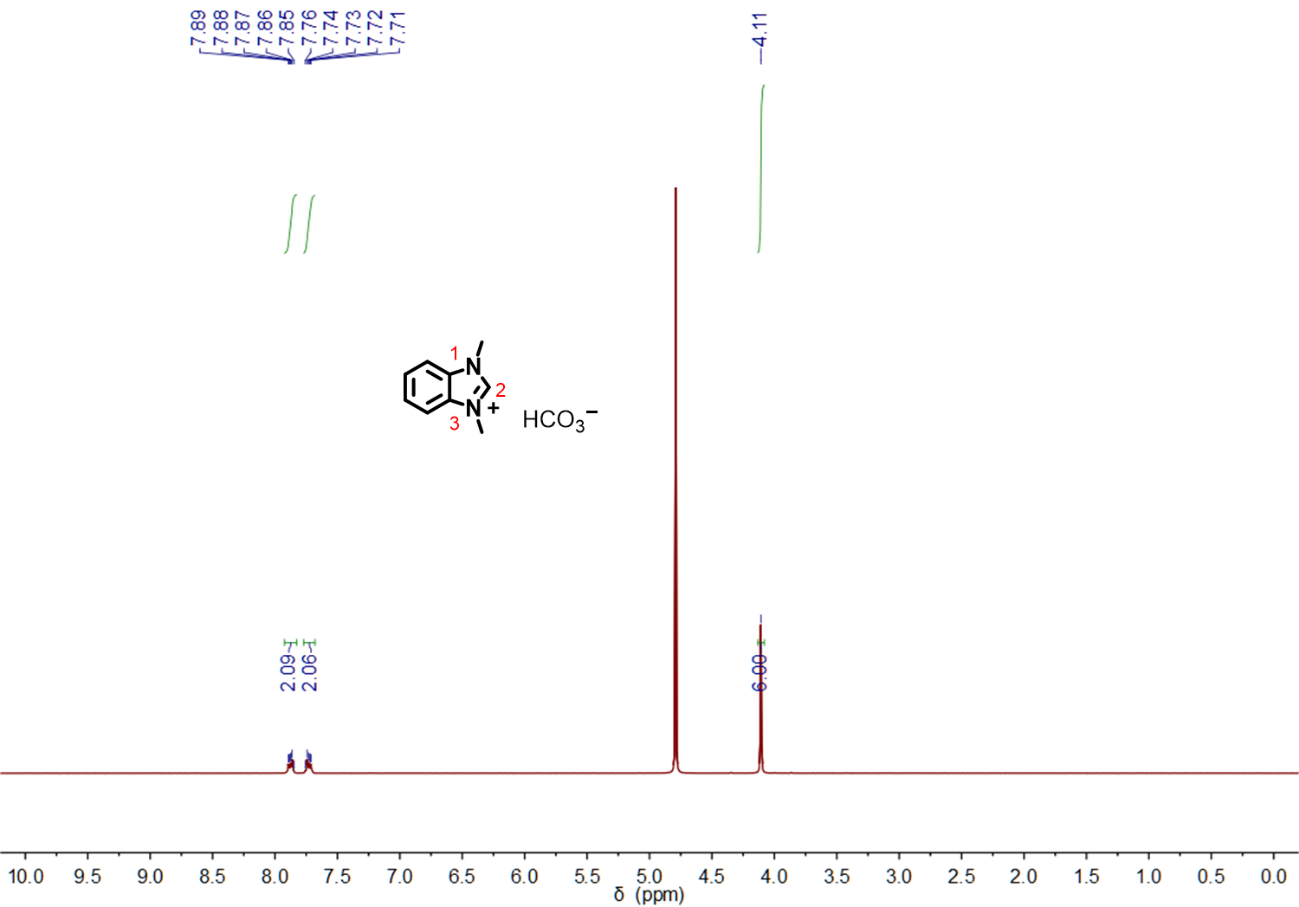}
    \caption{\textsuperscript{1}H NMR spectrum of \textbf{3} (NHC\textsuperscript{Me}, 300 MHz, D\textsubscript{2}O).}
    \label{fig:NMR_NHCMe}
\end{figure}

\begin{figure}
    \centering
    \includegraphics[width=0.8\linewidth]{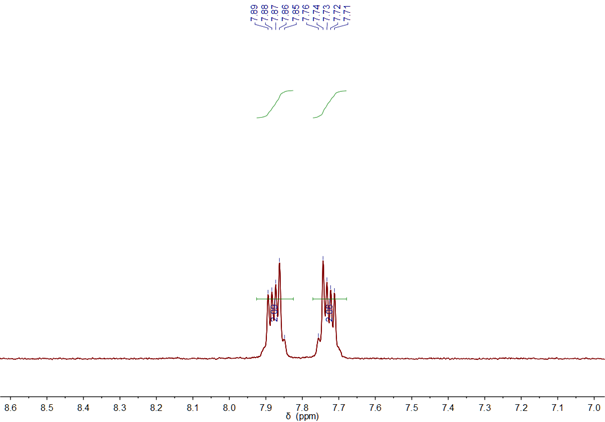}
    \caption{Expanded regions of the \textsuperscript{1}H NMR spectrum of \textbf{3} (NHC\textsuperscript{Me}, 300 MHz, D\textsubscript{2}O).}
    \label{fig:Closeup_NMR_NHCMe}
\end{figure}

\begin{figure}
    \centering
    \includegraphics[width=0.8\linewidth]{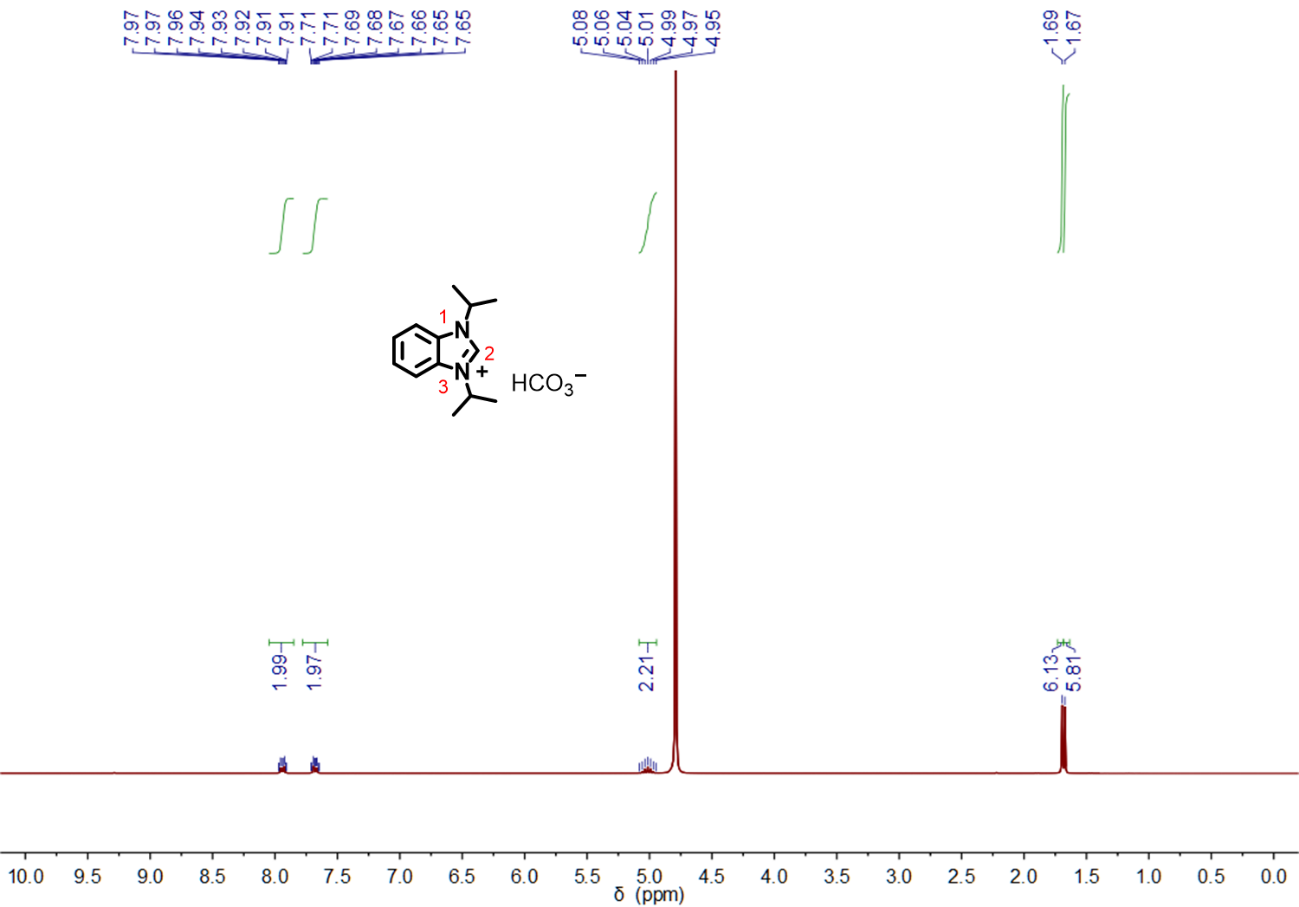}
    \caption{\textsuperscript{1}H NMR spectrum of \textbf{6} (NHC\textsuperscript{\textit{i}Pr}, 300 MHz, D\textsubscript{2}O).}
    \label{fig:NMR_NHCiPr}
\end{figure}

\begin{figure}
    \centering
    \includegraphics[width=0.8\linewidth]{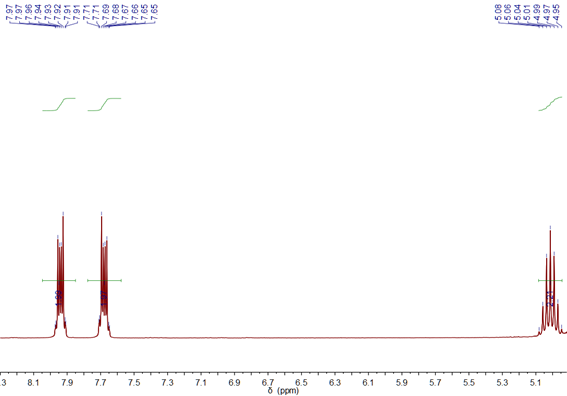}
    \caption{Expanded regions of the \textsuperscript{1}H NMR spectrum of \textbf{6} (NHC\textsuperscript{\textit{i}Pr}, 300 MHz, D\textsubscript{2}O).}
    \label{fig:Closeup_NMR_NHCiPr}
\end{figure}

\newpage

\section{Unit Cell Assignment of NHC SAMs on Au(111)}  

\begin{figure*}[h]
    \centering
    \includegraphics[width=0.7\textwidth]{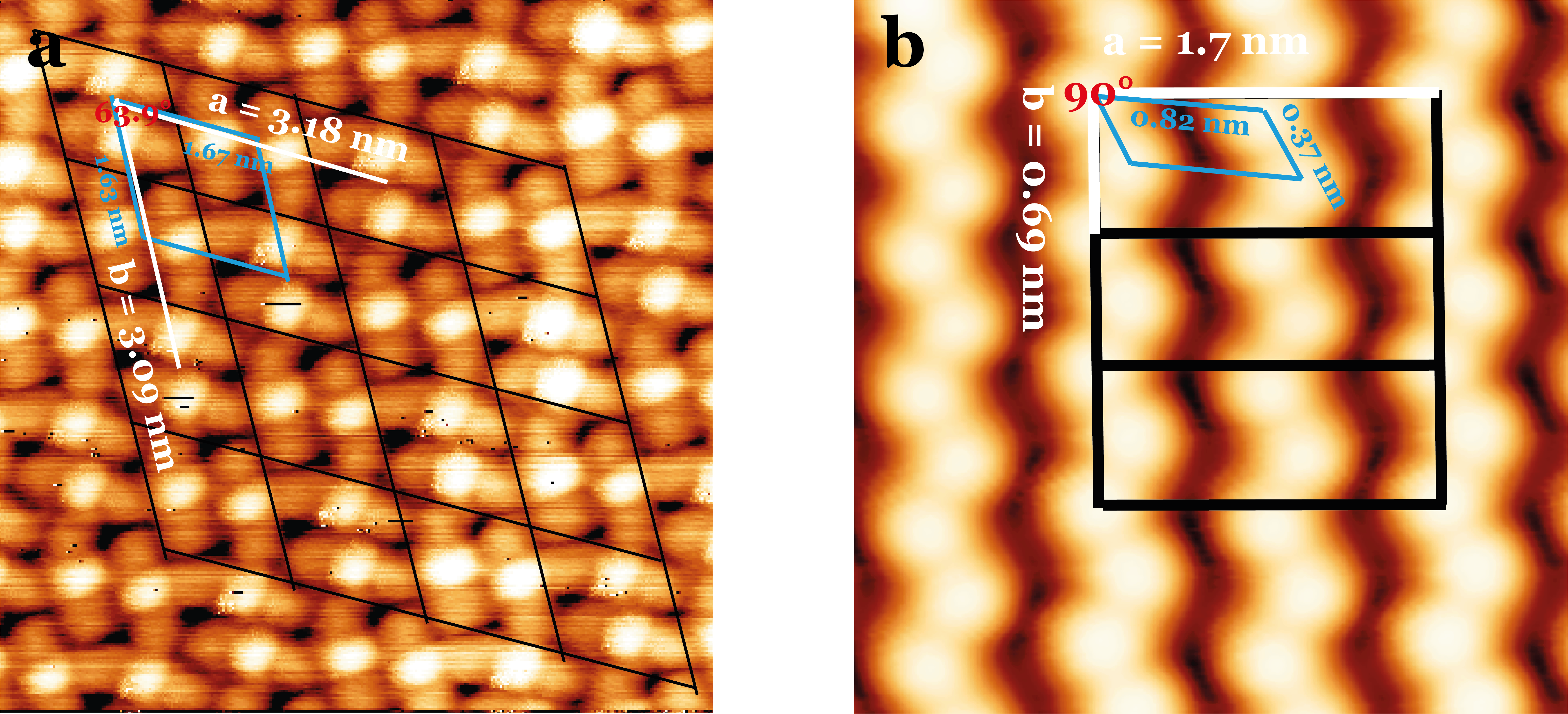}
    \caption{\textbf{a} STM-based unit cell assignment of the NHC\textsuperscript{Me} SAM on Au(111), showing a zigzag dimer packing with an oblique unit cell. \textbf{b} Unit cell assignment of the NHC\textsuperscript{\textit{i}Pr} SAM revealing a rectangular unit cell derived from alternating molecular rows.}
    \label{fig:SI_unitcells}
\end{figure*}

Figure~\ref{fig:SI_unitcells}a shows the structural characterization of self-assembled monolayers (SAMs) formed by NHC\textsuperscript{Me} and NHC\textsuperscript{\textit{i}Pr} on Au(111), based on high-resolution STM images. For NHC\textsuperscript{Me}, molecules assemble into a periodic zigzag pattern consisting of adatom-coordinated dimers. The resulting SAM forms an oblique unit cell with lattice parameters of 3.18~nm and 3.09~nm and an internal angle of 63.9$^\circ$. This structure reflects the alternating dimer motif and is observed to form multiple rotational domains across the surface.

In Figure~\ref{fig:SI_unitcells}b, the unit cell of the NHC\textsuperscript{\textit{i}Pr} self-assembled monolayer (SAM) is extracted from a region with well-resolved local order. The upright-standing carbene molecules organize into zigzagging rows that repeat every second row, resulting in a rectangular unit cell containing four NHC\textsuperscript{\textit{i}Pr} monomers. The lattice constants are measured as $a = 1.7$~nm and $b = 0.69$~nm, with an internal angle of 90$^\circ$, indicative of a staggered row-wise packing on Au(111). This arrangement is consistent with the known upright binding geometry of bulky NHC ligands via single gold adatoms.\cite{larrea2017, inayeh2021self}

\section{Additional observations on N-heterocyclic carbenes on Au(111)} 
\subsection*{Temperature-Dependent Desorption of NHC\textsuperscript{Me}}

\begin{figure}[h]
    \centering
    \includegraphics[width=\textwidth]{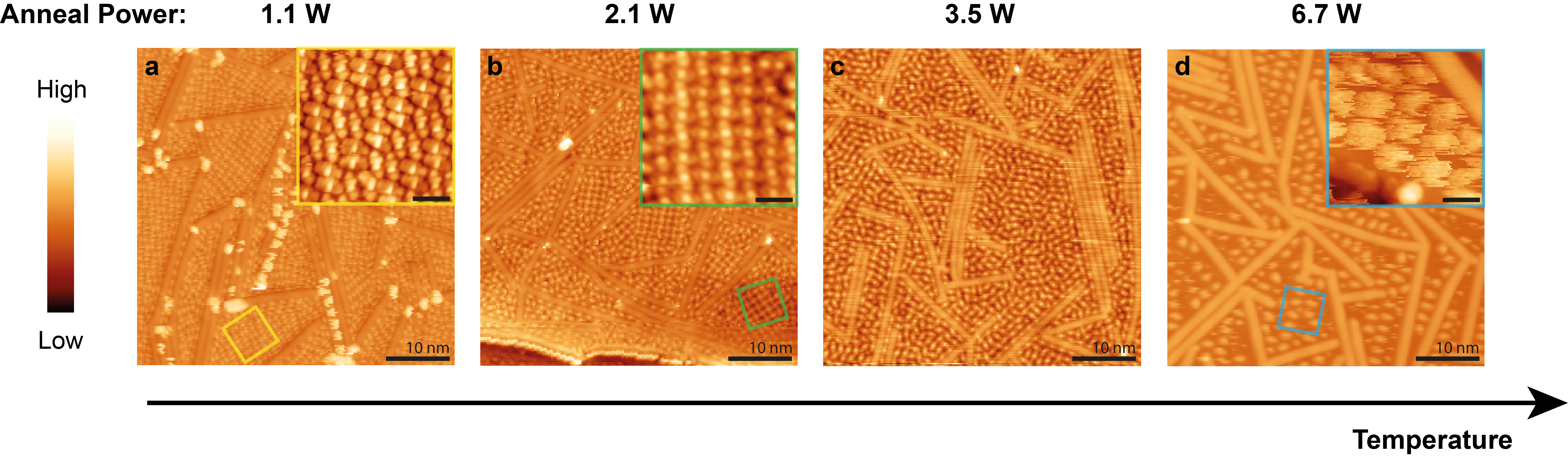}
    \caption{STM images showing temperature-dependent desorption of NHC\textsuperscript{Me} on Au(111) in the presence of 7-AGNRs. \textbf{a} Sample annealed at 1.1~W during deposition, showing an intact SAM and partial desorption of the weakly bound adlayer. \textbf{b} After annealing at 2.1~W for 1 hour, changes in the SAM structure indicate the onset of desorption within the monolayer regime. \textbf{c} Annealing at 3.5~W leads to further desorption and increased disorder among remaining molecules. \textbf{d} At 6.7~W (approx.\ 250$~^\circ\mathrm{C}$), residual NHC\textsuperscript{Me} remains on the surface but appears mobile. Complete desorption is typically achieved at 350$~^\circ\mathrm{C}$. Annealing power refers to the output of a resistive heating stage.}
    \label{fig:SI_annealing_desorption}
\end{figure}

Figure \ref{fig:SI_annealing_desorption}\textbf{a} shows a sample of NHC\textsuperscript{Me} with 7-AGNRs, annealed at 1.1~W (direct temperature measurement was not possible due to the pyrometer detection limit) during deposition using a resistive heating stage. The SAM remains intact, while most of the weakly bound adlayer desorbs, enabling quick and reliable sample preparation.  
Figure \ref{fig:SI_annealing_desorption}\textbf{b} shows the same sample annealed at 2.1~W for 1 hour. At this point, changes in the structure of the SAM are already visible, suggesting partial desorption of NHC\textsuperscript{Me} from the monolayer, with the remaining molecules rearranging to fill the gaps.  
In Figure \ref{fig:SI_annealing_desorption}\textbf{c}, the sample is annealed at 3.5~W, resulting in further desorption. The remaining molecules appear completely disordered.  
The applied powers refer to the output of the resistive heating stage (in watts). The annealing power of 6.7~W (Figure \ref{fig:SI_annealing_desorption}\textbf{d}) corresponds to approximately 250$~^\circ\mathrm{C}$, where some NHC\textsuperscript{Me} residues remain, though they appear mobile. Complete SAM desorption is reliably achieved by annealing at approximately 350$~^\circ\mathrm{C}$ for a few minutes. However, the precise desorption threshold has not been systematically studied.

\subsection*{Room Temperature Desorption of NHC\textsuperscript{Me}}

\begin{figure}[h]
    \centering
    \includegraphics[width=0.7\textwidth]{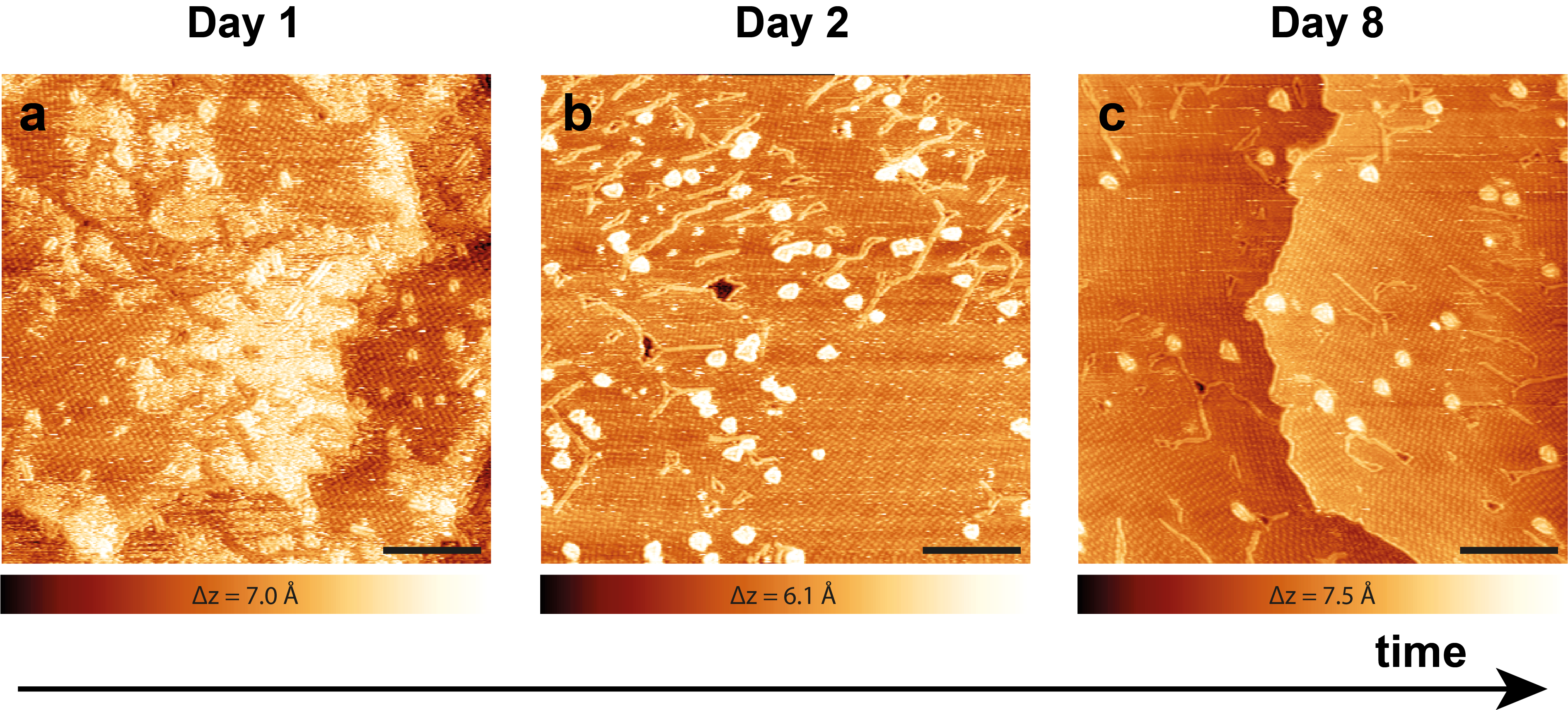}
    \caption{STM images showing the desorption behavior of NHC\textsuperscript{Me} over time at room temperature. \textbf{a} Sample with excess NHC\textsuperscript{Me} imaged directly after deposition. \textbf{b} Sample imaged 24 hours later, showing that the NHC\textsuperscript{Me} adlayer has largely desorbed, leaving behind a clean SAM. \textbf{c} Sample imaged after 8 days, demonstrating that the SAM remains stable in the monolayer regime. All STM measurements were conducted at room temperature. Scale Bar: 10~nm}
    \label{fig:SI_RT_desorption}
\end{figure}

Figure \ref{fig:SI_RT_desorption}\textbf{a} shows a SAM of NHC\textsuperscript{Me} prepared with an excess adlayer. As seen in the images, the adlayer gradually desorbs at room temperature, primarily within the first 24 hours. The excess material visible in panel \textbf{a} has completely disappeared in Figure \ref{fig:SI_RT_desorption}\textbf{b}, leaving behind a clean monolayer. After this initial period, the SAM remains stable under UHV conditions, as demonstrated in Figure \ref{fig:SI_RT_desorption}\textbf{c}, which shows the same sample after 8 days storage in UHV. These observations confirm that a well-defined SAM can be reliably obtained by depositing excess NHC\textsuperscript{Me} and allowing the weakly bound adlayer to desorb overnight under vacuum.

\subsection*{NHC\textsuperscript{Me} SAM Dimer Formation -- Surface Vacancies in the Au(111) Substrate}

\begin{figure}[h]
    \centering
    \includegraphics[width=0.3\textwidth]{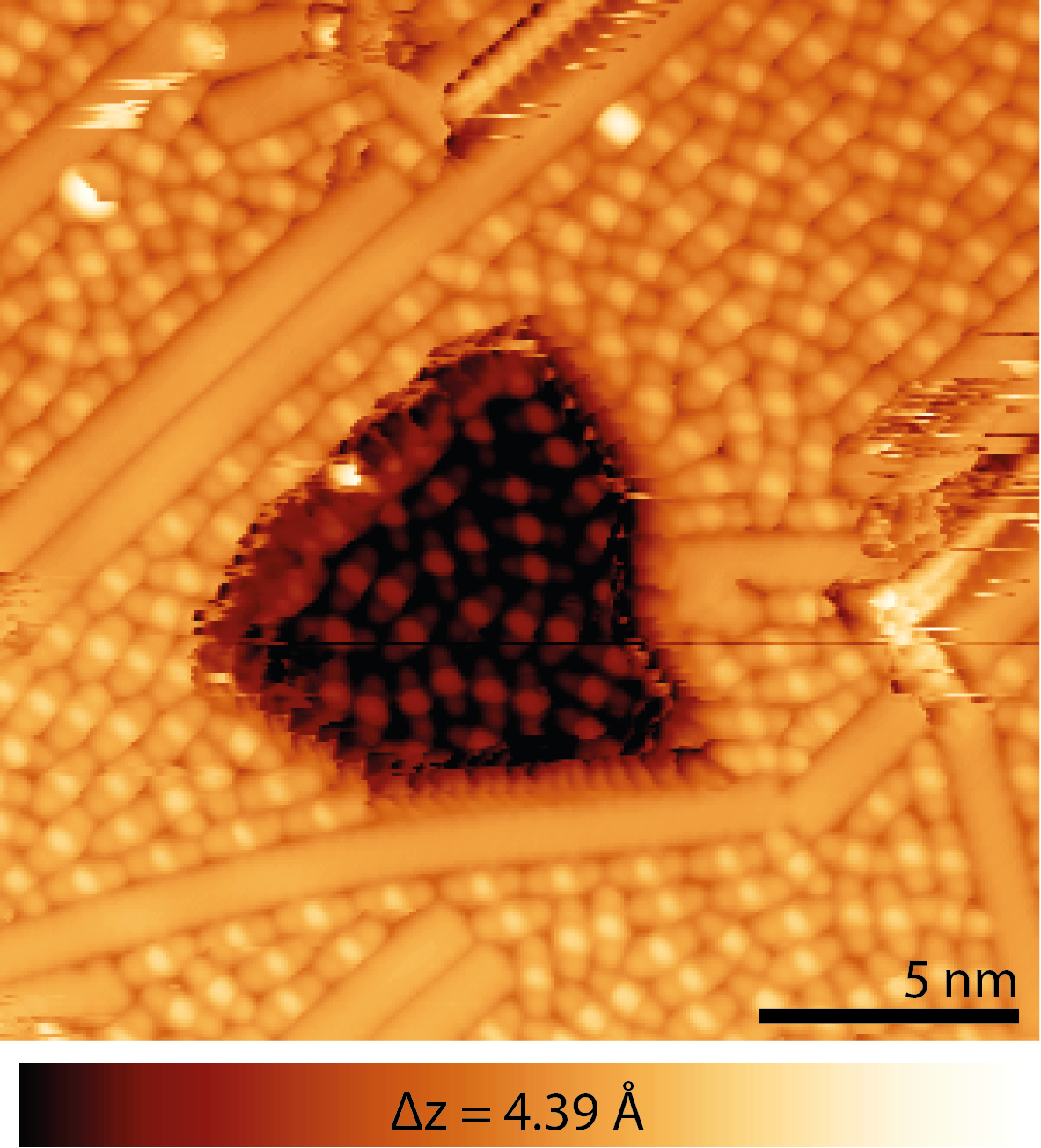}
    \caption{STM image of a NHC\textsuperscript{Me} SAM on Au(111), showing randomly 
    distributed surface vacancies in the Au(111) substrate. These features are attributed to the 
    extraction of Au atoms during dimer formation, which results in adatom-mediated coordination 
    of NHC\textsuperscript{Me} pairs and leaves behind pits in the Au(111) surface lattice.}
    \label{fig:SI_holes}
\end{figure}

Further evidence for the dimerization of NHC\textsuperscript{Me} is provided by the observation 
of surface vacancies in the Au(111) substrate beneath the SAM. These features are attributed to 
the extraction of individual Au atoms during the formation of NHC\textsuperscript{Me} dimers, 
which coordinate over adatoms. As shown in Figure~\ref{fig:SI_holes}, these vacancies appear at 
random locations on the surface and vary in size. Their presence is consistent with previous 
studies demonstrating that dimer formation is accompanied by Au adatom extraction, leaving 
behind vacancies in the Au(111) surface lattice.

\section{Statistical Analysis of Intercalation Yield}

To quantitatively assess the efficiency of GNR intercalation into the NHC\textsuperscript{Me} SAM, we performed a statistical analysis of segment lengths identified as either intercalated or embedded. Segment classification was carried out manually using STM image analysis tools, followed by length histogram generation.

\begin{figure}[h]
    \centering
    \includegraphics[width=\textwidth]{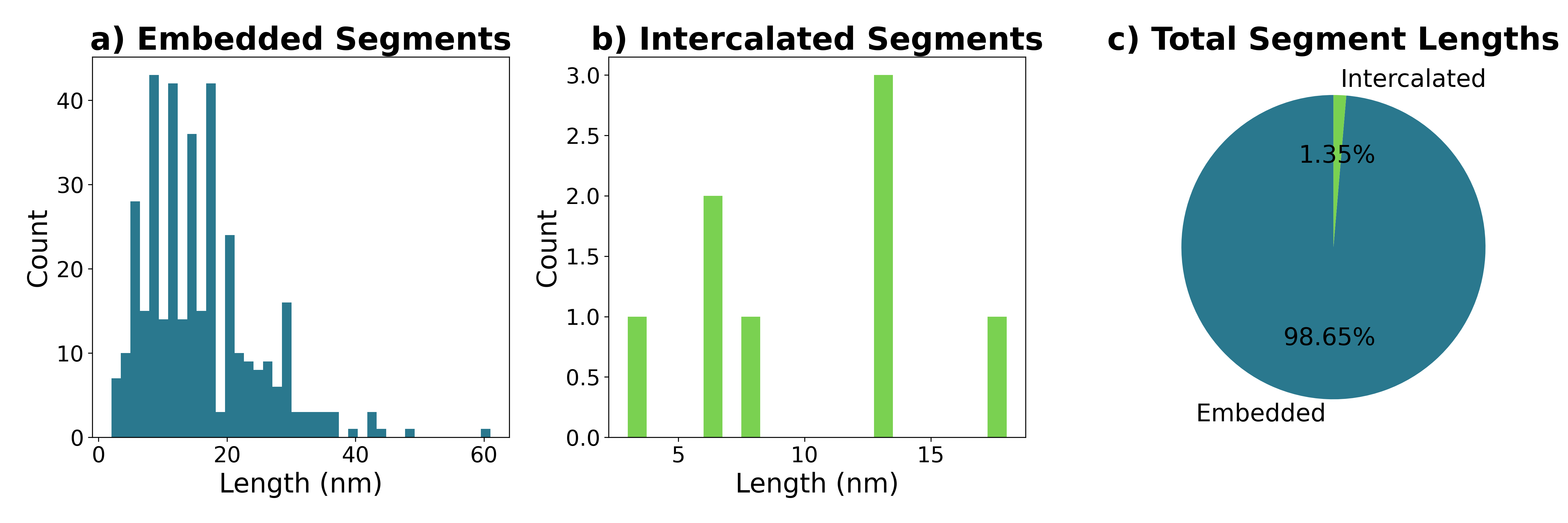}
    \caption{%
    \textbf{a} Histogram showing the length distribution of embedded 7-AGNR segments within the NHC\textsuperscript{Me} SAM. 
    \textbf{b} Histogram of intercalated 7-AGNR segment lengths under the same conditions. 
    \textbf{c} Relative contributions of embedded ($5.86 \times 10^{3}\,\mathrm{nm}$) and intercalated ($80\,\mathrm{nm}$) GNR segment lengths. The resulting intercalation yield is 1.35\%.
    }
    \label{fig:SI_Yield_Analysis}
\end{figure}

Figure~\ref{fig:SI_Yield_Analysis}\textbf{a}-\textbf{b} show histograms of the segment lengths for embedded and intercalated GNR portions, respectively. Embedded GNRs exhibit a broad length distribution with significantly higher frequency, reflecting their dominance in the sample. In contrast, intercalated GNRs are rare and predominantly short in length.

To quantify the intercalation yield, we calculated the total cumulative segment length in each category. The total embedded GNR length was determined to be $5.86 \times 10^{3}$~nm, while the total length of intercalated segments was $80$~nm. From these values, the intercalation yield is defined as  the fraction of total measured GNR length exhibiting intercalated appearance, yielding a value of 1.35\%. This low yield underscores the limited efficiency of NHC-assisted intercalation under the current preparation conditions.

A visual summary of this analysis is shown in Figure~\ref{fig:SI_Yield_Analysis}\textbf{c}, where the total lengths are represented as a stacked bar chart. This representation emphasizes the low contribution of intercalated ribbons compared to the embedded majority.

\section{Geometric optimization of different 7-AGNR/NHC configurations}

Table~\ref{tab:SI_calculations} summarizes the calculated relative energies for the various configurations of 7-AGNR with NHC\textsuperscript{Me} and NHC\textsuperscript{\textit{i}Pr}.  
All energies are given relative to the corresponding fully embedded reference configuration ($\Delta E = 0$~eV).  

For NHC\textsuperscript{\textit{i}Pr}, configurations with one or three NHC units placed below the GNR do not relax to stable intercalated geometries. In both cases, the system reverts toward the embedded configuration during geometry optimization. These configurations are therefore not included in the table and are not considered metastable states.

\begin{table}[h!]
\centering
\begin{tabular}{ccc}
\hline
\textbf{Motif} & \textbf{NHC\textsuperscript{Me} ($\Delta E$ [eV])} & \textbf{NHC\textsuperscript{\textit{i}Pr} ($\Delta E$ [eV])} \\ \hline
fully embedded & 0.00 & 0.00 \\
1 on top       & 1.74 & 1.96 \\
2 on top       & 3.69 & 4.02 \\
3 on top       & 6.13 & 6.26 \\
1 below        & 3.56 & -- \\
2 below        & 4.44 & 7.18 \\
3 below        & 4.53 & -- \\ 
\hline
\end{tabular}
\caption{Calculated relative energies ($\Delta E$) for different 7-AGNR/NHC configurations. Entries marked with ``--'' correspond to configurations that do not relax to stable intercalated geometries and are therefore not included as metastable states.}
\label{tab:SI_calculations}
\end{table}

\begin{figure}[h!]
    \centering
    \includegraphics[width=0.7\linewidth]{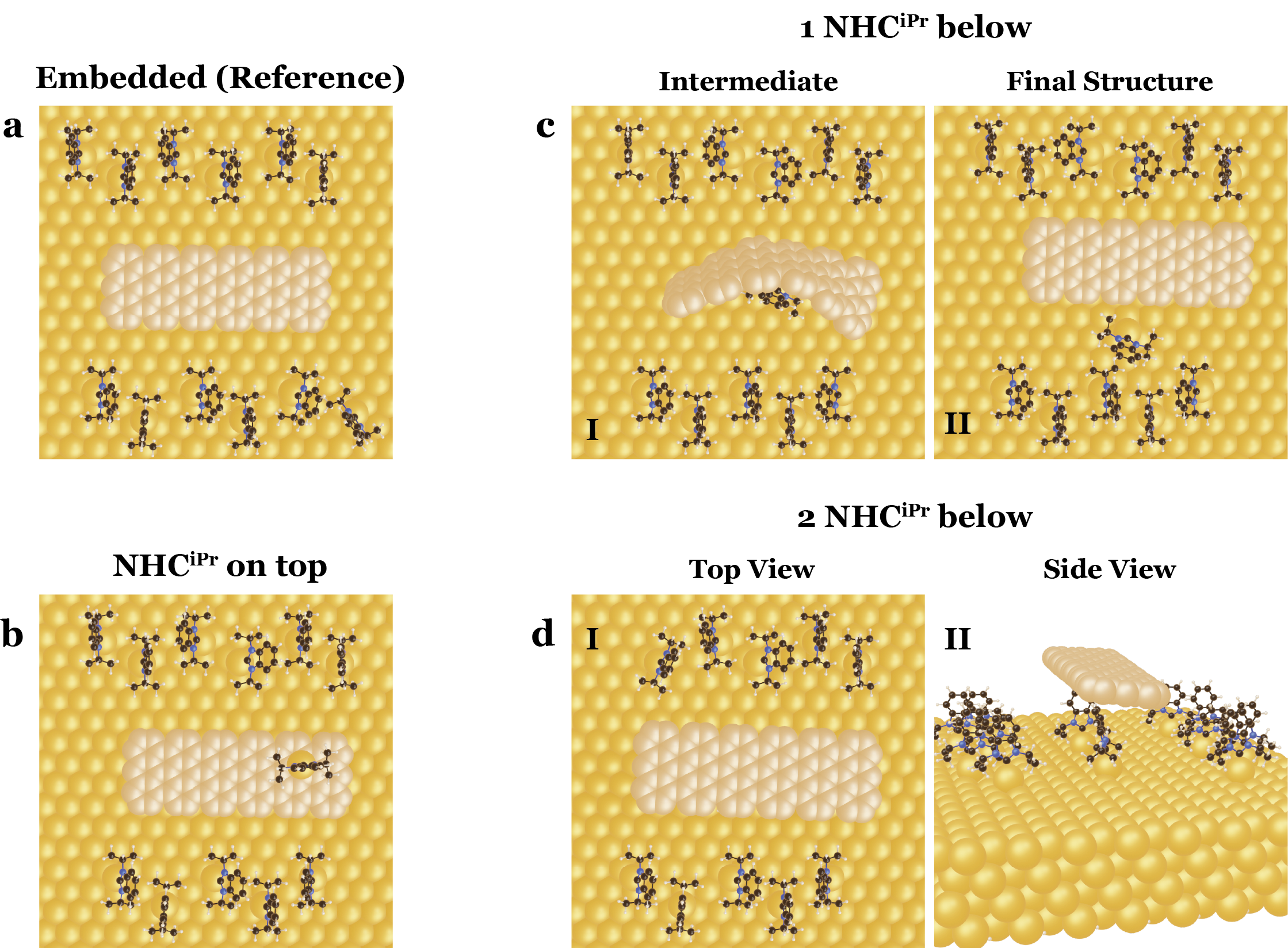}
    \caption{
    Geometrically optimized structures of 7-AGNR/NHC\textsuperscript{\textit{i}Pr} configurations.  
    \textbf{a} Embedded configuration of a 7-AGNR within 12 standing NHC\textsuperscript{\textit{i}Pr} monomers on Au(111), used as energetic reference.  
    \textbf{b} Configuration with one NHC\textsuperscript{\textit{i}Pr} placed on top of the GNR.  
    \textbf{c} (\textbf{I}) Intermediate configuration during the geometric optimization attempted intercalation of one NHC\textsuperscript{\textit{i}Pr} below the GNR; the ribbon  bends sideways towards the surface eventually pushing the NHC\textsuperscript{\textit{i}Pr} out from below, (\textbf{II}) Final optimized structure showing reversion to the embedded configuration,   
    \textbf{d} (\textbf{I}) Top view of the configuration with two NHC\textsuperscript{\textit{i}Pr} below the GNR after relaxation, where the ribbon rests on top of both NHCs, (\textbf{II}) Corresponding side view of (\textbf{I}), confirming the stabilized geometry on top of two NHC\textsuperscript{\textit{i}Pr} monomers.  
    }
    \label{fig:SI_NHCiPr_simulations}
\end{figure}

To illustrate this process, Figure~\ref{fig:SI_NHCiPr_simulations}\textbf{c-I} shows the intermediate state during relaxation, where the GNR bends almost perpendicular to the surface, displacing the NHC. The final optimized structure is shown in Figure~\ref{fig:SI_NHCiPr_simulations}\textbf{c-II}.  

For reference, Figure~\ref{fig:SI_NHCiPr_simulations}\textbf{a} shows the embedded configuration (energetic reference), Figure~\ref{fig:SI_NHCiPr_simulations}\textbf{b} shows the structure with one NHC\textsuperscript{\textit{i}Pr} on top of the GNR, and Figure~\ref{fig:SI_NHCiPr_simulations}\textbf{d} displays the configuration with two NHC\textsuperscript{\textit{i}Pr} below the ribbon.  
To aid visualization of the latter, the corresponding side view is provided in Figure~\ref{fig:SI_NHCiPr_simulations}\textbf{d-II}, confirming that the optimized geometry sits above the two NHC units.

\section{Temperature-Dependent Raman Spectroscopy of NHC\textsuperscript{\textit{i}Pr}-Functionalized 7-AGNRs}

\begin{figure}[h]
    \centering
    \includegraphics[width=\textwidth]{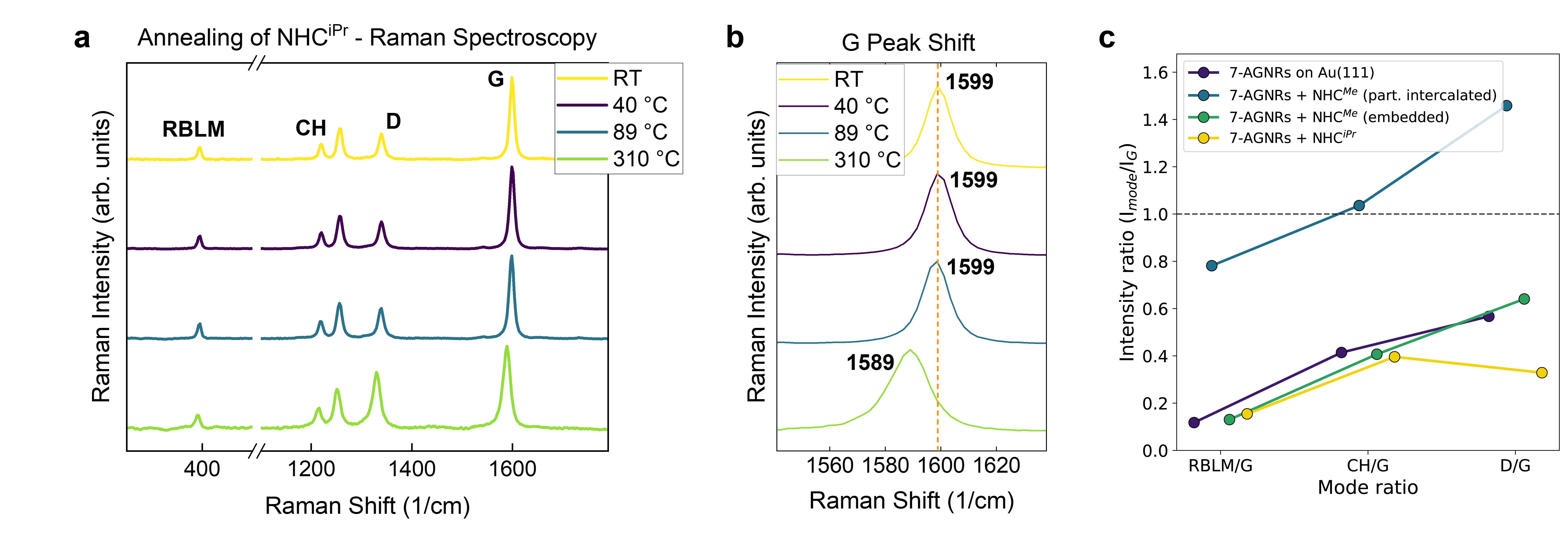}
    \caption{\textbf{a} Raman spectra of 7-AGNRs functionalized with NHC\textsuperscript{\textit{i}Pr} measured at four different temperatures (RT, 40 $^\circ$C, 89 $^\circ$C, and 310 $^\circ$C), showing consistent spectral shape up to 89 $^\circ$C and a clear downshift of the G-mode at 310 $^\circ$C. \textbf{b} G-mode peak position versus temperature. The G-mode remains constant at 1599~cm$^{-1}$ until desorption at 310 $^\circ$C, which induces a shift to 1589~cm$^{-1}$. All Raman spectra were acquired using a laser power of 20~mW and an integration time of 1~s.
    \textbf{c} Raman mode intensity ratios (RBLM/G, CH/G, and D/G) for all four sample configurations discussed in the main text. Only the partially intercalated NHC\textsuperscript{Me} sample exhibits ratios exceeding unity, while all other samples remain below one. All Raman spectra were acquired using a laser power of 20~mW and an integration time of 1~s.
    }
    \label{fig:SI_iPr_Temp_Raman}
\end{figure}

To evaluate the thermal stability and influence of NHC\textsuperscript{\textit{i}Pr} on the Raman response of 7-armchair graphene nanoribbons (7-AGNR), we conducted temperature-dependent Raman spectroscopy at four different sample temperatures: room temperature (RT), 40 $^\circ$C, 89 $^\circ$C, and 310 $^\circ$C. The resulting spectra are shown in Figure~\ref{fig:SI_iPr_Temp_Raman}\textbf{a}. These measurements allow us to assess both the desorption behavior of the carbene layer and potential structural or electronic changes induced upon thermal treatment.

The G-mode position remains largely unchanged from room temperature up to 89~$^\circ$C, consistently centered at approximately 1599~cm$^{-1}$. Upon heating to 310~$^\circ$C, a shift to 1589~cm$^{-1}$ is observed, as shown in Figure~\ref{fig:SI_iPr_Temp_Raman}\textbf{b}. Since this spectrum was recorded at elevated temperature, the observed redshift primarily reflects the well-established negative temperature coefficient of the graphene G mode, which arises from anharmonic phonon interactions and thermal expansion.\cite{guo2022phonon, bonini2007phonon} While desorption of NHC\textsuperscript{\textit{i}Pr} may occur at elevated temperatures, the Raman measurement at 310~$^\circ$C does not allow a direct correlation between the observed shift and carbene desorption. Therefore, the data are more consistently explained by thermal effects rather than by a change in the chemical environment.

In comparison, previous experiments with NHC\textsuperscript{Me} revealed the onset of desorption already around 50 $^\circ$C, as evidenced by exposure of embedded GNRs on the Au(111) surface. For NHC\textsuperscript{Me}, Raman shifts between embedded and intercalated GNRs were minimal, with both showing a slight upshift in G-mode frequency relative to GNRs directly on Au(111). In contrast, the NHC\textsuperscript{\textit{i}Pr}-coated sample does not exhibit any upshift in the G-mode prior to desorption. Furthermore, even after desorption at 310 $^\circ$C, the resulting spectrum remains consistent with that of the pristine 7-AGNR on Au(111), indicating that the interaction of NHC\textsuperscript{\textit{i}Pr} with the GNR/Au interface differs from that of NHC\textsuperscript{Me}.

This behavior suggests that the spectroscopic influence of NHC\textsuperscript{\textit{i}Pr} on GNRs is weaker or qualitatively different, possibly due to its upright binding geometry or reduced surface coverage. Unlike NHC\textsuperscript{Me}, which adsorbs in a flat-lying dimeric configuration and interacts more directly with both the substrate and the GNRs, NHC\textsuperscript{\textit{i}Pr} may induce less charge redistribution or structural strain, resulting in a negligible Raman shift. The absence of any spectral change prior to desorption further confirms the lack of intercalation in this system, consistent with STM observations.

Figure~\ref{fig:SI_iPr_Temp_Raman}\textbf{c} summarizes the Raman mode intensity ratios (RBLM/G, CH/G, and D/G) for all four sample configurations discussed in the main text.
Only the partially intercalated NHC\textsuperscript{Me} sample exhibits intensity ratios exceeding unity, whereas all other configurations (including NHC\textsuperscript{\textit{i}Pr}) remain below one.
This further supports the conclusion that NHC\textsuperscript{\textit{i}Pr} does not intercalate the 7-AGNRs, consistent with both STM observations and the absence of spectroscopic signatures of decoupling.

\end{document}